\documentclass[conference,compsoc]{IEEEtran}
\ifCLASSOPTIONcompsoc
  \usepackage[nocompress]{cite}
\else
  \usepackage{cite}
\fi
\ifCLASSINFOpdf
\else
\fi

\usepackage{epsfig,endnotes}
\usepackage{url}
\usepackage{xcolor}
\usepackage{minted}
\usepackage{caption}
\usepackage{rotating}
\usepackage{booktabs,tabularx,array,multirow,multicol,colortbl}
\usepackage{pifont}
\usepackage[most]{tcolorbox}
\usepackage{enumitem}
\usepackage{fontawesome5}
\usepackage{tikz}
\usepackage{hyperref}

\usepackage{placeins}

\newcommand*\emptycirc[1][0.72ex]{\tikz\draw[thick] (0,0) circle (#1);} 
\newcommand*\halfcirc[1][0.8ex]{%
  \begin{tikzpicture}
  \draw[fill] (0,0)-- (90:#1) arc (90:270:#1) -- cycle ;
  \draw[thick] (0,0) circle (#1);
  \end{tikzpicture}}
\newcommand*\fullcirc[1][0.8ex]{\tikz\fill (0,0) circle (#1);}

\definecolor{mylightblue}{HTML}{D9EAFD}

\newcommand{\mypar}[1]{\smallskip\noindent\textit{\textbf{#1.}}}

\newcounter{vksec}[subsection] % Reset vsec at each new section
\renewcommand{\thevksec}{\arabic{vksec}} % 1, 2, 3, ...

\newcommand{\vksec}[1]{%
    \refstepcounter{vksec}%
    \subsubsection*{KV.\thevksec\space - #1}%
}

\newcounter{vssec}[subsection] % Reset vsec at each new section
\renewcommand{\thevssec}{\arabic{vssec}} % 1, 2, 3, ...

\newcommand{\vssec}[1]{%
    \refstepcounter{vssec}%
    \subsubsection*{SV.\thevssec\space - #1}%
}

\newcommand{\secanf}[1]{%
  \subsubsection*{SA.F\space - #1}\text{\newline}%
}

\newcommand{\secanh}[1]{%
    \subsubsection*{SA.H\space - #1}\text{\newline}%
}

\newcommand{\mseff}[1]{%
    \subsubsection*{MS\space - #1}\text{\newline}%
}

\newcounter{upsec}[subsection] % Reset vsec at each new section
\renewcommand{\theupsec}{\arabic{upsec}} % 1, 2, 3, ...

\newcommand{\upsec}[1]{%
    \refstepcounter{upsec}%
    \subsubsection*{UP.\theupsec\space - #1}%
}

\newtcolorbox[auto counter, number within=section]{marcosboxwhite}{%
  colback=gray!0, % Background color
  colframe=black, % Border color
  toprule=-0.25mm, % Remove top border
  bottomrule=-0.25mm, % Remove bottom border
  leftrule=1mm, % Add left border
  rightrule=-0.25mm, % Remove right border
  left=0.1mm, right=0.1mm, top=0.1mm, bottom=0.1mm, % Padding
  enhanced,
  arc=0mm,
  breakable = true,
}

\newtcolorbox[auto counter, number within=section]{marcosboxgreyopen}{%
  colback=gray!10, % Background color
  colframe=black, % Border color
  toprule=-0.25mm, % Remove top border
  bottomrule=-0.25mm, % Remove bottom border
  leftrule=1mm, % Add left border
  rightrule=-0.25mm, % Remove right border
  left=0.1mm, right=0.1mm, top=0.1mm, bottom=0.1mm, % Padding
  enhanced,
  arc=0mm,
  breakable = true,
}

\newtcolorbox[auto counter, number within=section]{marcosboxgreyclosed}{%
  colback=gray!10, % Background color
  colframe=black, % Border color
  toprule=-0.25mm, % Remove top border
  bottomrule=-0.25mm, % Remove bottom border
  leftrule=-0.25mm, % Add left border
  rightrule=1mm, % Remove right border
  left=0.1mm, right=0.1mm, top=0.1mm, bottom=0.1mm, % Padding
  enhanced,
  arc=0mm,
  breakable = true,
}

\usepackage[nolist]{acronym}
\begin{acronym}
    \acro{PoC}{proof-of-concept}
    \acro{ML}{Machine Learning}
    \acro{AI}{Artificial Intelligence}
    \acro{DL}{Deep Learning}
    \acro{LLM}{Large Language Model}
\end{acronym}

\usepackage{fancyhdr}
\fancypagestyle{firstpagestyle}{
  \fancyhf{}
  \fancyhead[C]{\textbf{(to appear in) 2026 IEEE Symposium on Security and Privacy (SP)}}
  
}

\begin{document}

% \title{Secure by Name, Vulnerable by Design: \\
% On the (In)Security of Machine Learning Model Sharing}
\title{
% Secure by Name, Vulnerable by Design: \\
On the (In)Security of Loading Machine Learning Models
% On the (In)Security of Machine Learning Model Sharing\\
% {\color{blue}On the (In)Security of Machine Learning Model Artifact Sharing} \\
% {\color{green}On the (In)Security of Shared Machine Learning Models} \\ 
% {\color{pink}On the (In)Security of Machine Learning Model Sharing Ecosystems} \\
% {\color{orange}On the (In)Security of Machine Learning Model Sharing:\\ An Ecosystem-Level Analysis}
}

\author{
\IEEEauthorblockN{
Gabriele Digregorio,
Marco Di Gennaro,
Stefano Zanero,
Stefano Longari,
Michele Carminati}
\IEEEauthorblockA{
Politecnico di Milano\\
Milan, Italy\\
\{gabriele.digregorio, marco.digennaro, stefano.zanero, stefano.longari, michele.carminati\}@polimi.it}
}


% compsoc conferences compared to traditional conferences), use this
%Telephone: (800) 555--1212, Fax: (888) 555--1212}


\maketitle

\thispagestyle{firstpagestyle} % Header for preprint

% Despite their benefits, these tools expose users to underexplored security risks, while security awareness remains limited among both practitioners and developers. 
\begin{abstract}
The rise of model sharing through frameworks and dedicated hubs makes \acl{ML} significantly more accessible. 
Despite its benefits, loading shared models exposes users to underexplored security risks, while security awareness remains limited among both practitioners and developers. 
To enable a more security-conscious approach in \acl{ML} model sharing, in this paper, we evaluate the security posture of frameworks and hubs, assess whether security-oriented mechanisms offer real protection, and survey how users perceive the security narratives surrounding model sharing.
Our evaluation shows that most frameworks and hubs address security risks partially at best, often by shifting responsibility to the user.
More concerningly, our analysis of frameworks advertising security-oriented settings and complete model sharing uncovered multiple 0-day vulnerabilities enabling arbitrary code execution.
Through this analysis, we show that, despite the recent narrative, securely loading \acl{ML} models is far from being a solved problem and cannot be guaranteed by the file format used for sharing.
Our survey shows that the security narrative leads users to consider security-oriented settings as trustworthy, despite the weaknesses shown in this work.
From this, we derive suggestions to strengthen the security of model-sharing ecosystems.
\end{abstract}


% \fi
\IEEEpeerreviewmaketitle

\section{Introduction}
\label{sec:intro}

% In recent years, \acf{ML} has evolved into a ubiquitous and transformative force across countless sectors. This widespread adoption reflects what many now describe as the \textit{democratization} of \ac{ML}: a growing trend in which advanced \ac{ML} tools and capabilities are no longer confined to specialized experts but are increasingly accessible to a broader audience.

% In recent years, the adoption of \acf{ML} has grown rapidly across diverse sectors. This growth is reflected in the development of advanced tools that were once the domain of experts and are now increasingly accessible to a broad audience. 
In recent years, the adoption of \acf{ML} has grown rapidly, with advanced tools once limited to experts now accessible to a broad audience.
Among other factors, this trend is driven by the rise of platforms for sharing pre-trained models~\cite{HuggingFaceHub, KaggleModels, kerashub2024, TensorFlowHub, PyTorchHub}. Such models are often advertised as ready-to-use, lowering entry barriers and enabling practitioners with different levels of expertise to incorporate \ac{ML} into their workflows.

This trend reflects the evolution of traditional software development, where open-source repositories enable the reuse and adaptation of algorithms, code, and libraries. Today, models are shared and reused, enabling a collaborative ecosystem that is reshaping how \ac{ML} is practiced. However, unlike the well-studied risks of code sharing and software supply chains~\cite{duan2021towards,ohm2020backstabber,ladisa2023sok,jiang2022empirical,casey2024large,shenao2025sok,meiklejohn2025machine}, the security implications of \ac{ML} model sharing remain largely underexplored. 
% Prior work on malicious code injection in models~\cite{hua2024malmodel,wang2021evilmodel,wang2022evilmodel2,zhao2024models} and on hub security gaps~\cite{jiang2023exploring,jiang2023empirical,zhao2024models,Jones2024What} is fragmented, resulting in limited awareness among users, framework developers, and sharing hubs. 
% The immaturity of the model sharing security field is further illustrated by Zhu et al.~\cite{zhu2025my}, who show how TensorFlow APIs can be abused for file and network access at inference time. 
Prior work on malicious code injection in models~\cite{hua2024malmodel,wang2021evilmodel,wang2022evilmodel2,zhao2024models} and on hub security gaps~\cite{jiang2023exploring,jiang2023empirical,zhao2024models,Jones2024What} remains fragmented, which may contribute to limited awareness among users, framework developers, and sharing hubs.
Zhu et al.~\cite{zhu2025my} further suggest open challenges in this space by showing that TensorFlow APIs can be abused for file and network access at inference time.
The growing relevance of the problem is also reflected in a recent DEF CON 33 talk~\cite{ParzianDefcon33}, which highlighted persistent risks in model sharing, such as insecure pickle deserialization.%~\footnote{Due to space constraints, we refer the interested reader to Appendix~\ref{appendix:related} for the complete review of related work.}.% and execution of raw \texttt{.exe} formats.

% This work builds on that observation, critically evaluating how \acl{ML} models are shared in practice, in terms of the mechanisms provided by popular \acl{ML} frameworks, approaches adopted by model hubs, and perception of the users. Our investigation is guided by the following research questions:

% Our work addresses the fragmented and immature state of security research on model sharing, aiming to raise awareness among everyday \ac{ML} users of its hidden risks, debunk common misconceptions, and encourage the security community to study these threats with the same rigor applied to open-source software at large.

% {\color{orange} Our work provides a systematic analysis of the security landscape of \ac{ML} model sharing. We assess major frameworks and sharing hubs, discover new vulnerabilities, and highlight a misalignment between perceived and actual security mechanisms, thereby underscoring the need for the same level of rigor in studying model-sharing threats as that applied to open-source software at large.}
Our work provides a systematic analysis of the security landscape of \ac{ML} model sharing. We evaluate major \ac{ML} frameworks and sharing hubs, uncover previously unknown vulnerabilities, and expose a misalignment between perceived and actual security mechanisms. Overall, our findings emphasize the need to apply the same level of rigor to the study of model-sharing threats as is currently devoted to open-source software security at large.

%We consider a threat model where attackers compromise the victim’s system by leveraging malicious model artifacts to achieve arbitrary code execution at load time.
We consider a threat model where attackers craft malicious model artifacts to compromise a victim’s system, with arbitrary code execution as the goal.
%The paper proceeds in three steps. First, we analyze the security posture of model-sharing solutions adopted by popular \ac{ML} frameworks as well as the distribution strategies employed by model hubs. Second, we assess the effectiveness of those solutions that are explicitly presented as security-oriented, extending this claim to the sharing of complete model objects rather than a limited scope, thereby promoting the narrative that a definitive solution to the model sharing problem has nearly been achieved. Third, we assess \ac{ML} users’ perceptions of these security postures.
% To formalize these three steps, we define the following research questions, which guide this work.
%{\color{red}This paper is structured in three steps, which are guided by the research questions defined below.}
Our research is driven by the questions defined below:
% The paper, in its structure, is driven by the research questions defined below.
% \begin{quote}
%     \textbf{RQ1}. What model-sharing mechanisms do the most widely used \ac{ML} frameworks and hubs offer, and what is the security posture of such approaches?
% \end{quote}
% \begin{quote}
%     \textbf{RQ2}. Are the security-oriented model-sharing approaches actually secure in practice?
% \end{quote}
% \begin{quote}
%     \textbf{RQ3}. Is the user's perception of the security posture consistent with reality, or does the security narrative affect their understanding of the risks?
% \end{quote}

\smallskip
%\mypar{RQ1} What model-sharing mechanisms do the most widely used \ac{ML} frameworks and hubs offer, and what is the security posture of such approaches?
\mypar{RQ1} \textit{What is the security posture of the model-sharing mechanisms adopted by popular \ac{ML} frameworks and hubs?}

\mypar{RQ2} \textit{Are approaches claiming security, while offering full model object sharing, actually secure?}
% \mypar{RQ2} \textit{Are approaches claiming security guarantees, while offering full model object sharing, actually secure?}

\mypar{RQ3} \textit{Is the user's perception of the security posture consistent with reality, or does the security narrative affect their understanding of the sharing-associated risks?
}
\smallskip

%classify framework-level sharing methods on two dimensions: distinguishing between code-based formats (e.g., pickle), which can include executable logic, and data-based formats (e.g., JSON), which typically store model structures or parameters descriptively; and on whether they produce what we define as \textit{self-contained} artifacts (i.e., including all components to restore a model) or rely on external dependencies. 

% To answer RQ1, we analyze the security of the most popular \ac{ML} frameworks and hubs, focusing on their sharing approaches and characteristics.
% We observe that while some frameworks do not address security at all, others provide security-oriented mechanisms based on shifting responsibility to other parts of the sharing workflow or by restricting the expressiveness of the model representation. Some others go further, providing complete model-sharing capabilities while explicitly promoting them as security-oriented. This narrative is reinforced through documentation and naming choices (e.g., Keras's ``safe mode''~\cite{KerasSavingLoading}), and is often justified by the use of data-based formats, assumed to inherently reduce the risk of arbitrary code execution.  Similarly, we examine the distribution mechanisms of popular model-sharing hubs, highlighting which implement safeguards, such as content scanning.

To answer RQ1, we analyze how the most popular \ac{ML} frameworks and hubs address model sharing. % and characteristics.
% To answer RQ1, we analyze the security of the most popular \ac{ML} frameworks and hubs, focusing on their sharing-related approaches. % and characteristics.
We observe that while some of them do not address security at all, others provide security-oriented mechanisms based on shifting responsibility to other parts of the sharing workflow or on restricting the expressiveness of the model representation. 
% Some frameworks go further, providing complete model-sharing capabilities while explicitly promoting themselves as security-oriented, whereas some hubs emphasize safeguards such as content scanning.
Some go further, with certain frameworks providing complete model-sharing capabilities while explicitly promoting themselves as security-oriented, and some hubs emphasizing the same security focus through safeguards such as content scanning.
This narrative is reinforced through documentation and naming choices (e.g., Keras's ``safe mode''~\cite{KerasSavingLoading}), and is often justified by the use of data-based formats, assumed to reduce the risk of arbitrary code execution.
This class of security-oriented sharing frameworks and hubs led to RQ2, which motivates a vulnerability assessment of their features.
In our investigation, we discovered a remarkably low number of pre-existing \textit{Common Vulnerabilities and Exposures} (referred to as \textit{CVEs}) affecting these mechanisms.
Through a manual reverse engineering analysis, we uncovered six 0-day vulnerabilities (i.e., previously undisclosed), each enabling arbitrary code execution.
Among these, we identified the first officially recognized CVEs targeting Keras's ``safe mode''~\cite{CVE-2025-1550, CVE-2025-8747}.
Collectively, these vulnerabilities challenge the widespread assumption that data-based formats (e.g., JSON) are inherently \textit{secure} when used to share full model objects. Overall, our findings in response to RQ2 reveal a critical gap between the \textit{security narrative} and the actual implementation.

% This understanding led to RQ3. To answer it, we conducted a survey targeting \ac{ML} practitioners to understand how frameworks and hubs' security narratives influence user perception. Results worryingly indicate that design terminology and a security-oriented claiming documentation significantly shape users' sense of security.
This understanding led to RQ3. To address it, we conducted a survey targeting \ac{ML} practitioners to examine how the security narratives promoted by frameworks and hubs influence user perception. 
% The results worryingly indicate that design terminology and security-oriented claims in documentation significantly shape users' sense of security.
The results worryingly indicate that security-oriented terminology and claims in documentation significantly shape users' sense of security.

We conclude our paper with takeaways and suggestions for users and developers, synthesized from the observations and insights obtained through our research questions.
%Results indicate that design terminology and documentation significantly shape users’ sense of security. This is concerning, especially given that we demonstrated these assurances do not always hold in practice.

% \mypar{Contributions} Summarizing, in this paper, we make the following contributions:

\mypar{Contributions} We make the following contributions: 
\begin{itemize}[left=0em, topsep=2.5pt]
    \item We systematize and analyze the security of \ac{ML} model-sharing mechanisms, covering both framework-level and hub-level perspectives.
    
    \item We assess methods that claim security while offering full model object sharing, identifying several CVEs, and challenging the common assumption that data-based formats are inherently secure.
    % \item We assess methods that claim security guarantees while offering full model object sharing, identifying several CVEs, and challenging the common assumption that data-based formats are inherently secure.

    \item We reveal a disconnect between the perceived security narrative (and the resulting community belief) and the reality through a survey on model sharing. 

    \item We provide takeaways for the community to promote a more security-aware culture in model sharing.
    % \item We provide takeaways and suggestions for the community, to promote a more security-aware culture in model sharing.
    
\end{itemize}

\mypar{Open Science} We provide all artifacts necessary to reproduce our empirical results. A repository, available at \url{https://github.com/necst/security-model-sharing}, contains the frozen version of our \ac{PoC} exploits, model artifacts, survey data, and analysis scripts.
\newcolumntype{L}[1]{>{\raggedright\arraybackslash}m{#1}}
% \newcolumntype{C}[1]{>{\centering\arraybackslash}m{#1}}   % centered, vertically centered

% % Convenience macros
% \newcommand{\cmark}{\textcolor{green!60!black}{\ding{51}}}        % green check
% \newcommand{\xmark}{\textcolor{red}{\ding{55}}}                   % red cross
% \newcommand{narrative}{\textcolor{yellow!80!black}{\Large\textasciitilde}} % big yellow tilde

% % Legend-only, baseline-aligned symbols
% % Legend-only symbols (smaller + baseline aligned)
% \newcommand{\legendcmark}{\textcolor{green!60!black}{\ding{51}}}
% \newcommand{\legendxmark}{\textcolor{red}{\ding{55}}}
% \newcommand{\legendpmark}{\raisebox{-0.35ex}{\textcolor{yellow!80!black}{\Large\textasciitilde}}}

% Convenience macros
%\newcommand{\cmark}{{\ding{51}}}        % green check
%\newcommand{\xmark}{{\ding{55}}}                   % red cross
%\newcommand{\pmark}{{\Large\textasciitilde}} % big yellow tilde

\newcommand{\cmark}{\scalebox{0.7}{\faCheck}}        
\newcommand{\xmark}{\scalebox{0.8}{\faTimes}}       
\newcommand{\pmark}{\raisebox{-0.35ex}{\scalebox{1.9}\textasciitilde}} % big yellow tilde

% Legend-only, baseline-aligned symbols
% Legend-only symbols (smaller + baseline aligned)
\newcommand{\legendcmark}{\cmark}
\newcommand{\legendxmark}{\xmark}
\newcommand{\legendpmark}{\pmark}

\newcommand{\supported}{{\fullcirc}}
\newcommand{\partiallysupported}{{\halfcirc}}
\newcommand{\notsupported}{{\textbf{\emptycirc}}}

\begin{table*}[th]
\centering
\small
\setlength{\tabcolsep}{5pt}
\renewcommand{\arraystretch}{1.2}

% \caption{Framework-level sharing: file formats, self-containment (weights and model), and security considerations (as stated in the official documentation and maintainer narrative).}
\caption{Framework-level ML model sharing mechanisms and security considerations, summarizing the persistence formats documented by the analyzed ML frameworks and indicating whether they are data- or code-based, self-contained (i.e., sufficient to reconstruct the full model), and their associated security models.
% \marco{protobuf + (and) raw data + (and) assets, chiarire dato che lui dice "I assume you mean "protobuf raw data assets", rather than 3 separate lines - the visual styling of the table should reflect that these are all just one thing."}
% \marco{"Green highlighting not clearly defined (same issue with Table 2)". Table caption could be more expressive (even more important for Table 2).}
% Framework-level sharing: documented characteristics and security considerations.
}
\label{tab:framework-level-sharing}
\resizebox{\linewidth}{!}{%
\begin{tabular}{@{}%
l  % Framework
l  % Setting
c  % File Format - Data
c  % File Format - Code
c  % Weights
c  % Model
c  % Claim
l  % Comments
@{}}
\toprule
\multicolumn{1}{l}{\textbf{Framework}} &
\multicolumn{1}{l}{\textbf{Setting}} &
\multicolumn{2}{l}{\textbf{File Format}} &
\multicolumn{2}{l}{\textbf{Self-contained}} &
\multicolumn{2}{l}{\textbf{Security model (security posture + commentary)}} \\
\cmidrule(lr){3-4}\cmidrule(lr){5-6}\cmidrule(lr){7-8}
&  & 
\multicolumn{1}{l}{\textbf{Data-based}} &
\multicolumn{1}{l}{\textbf{Code-based}} &
\multicolumn{1}{l}{\textbf{Weights}} &
\multicolumn{1}{l}{\textbf{Model}\textsuperscript{$\star$}} &
\multicolumn{1}{l}{\textbf{SO}} &
\multicolumn{1}{l}{\textbf{Comments}} \\
\midrule
% --- KERAS BLOCK ---
\multirow[c]{4}{*}[-0.6ex]{Keras~\cite{chollet2015keras, KerasSavingLoading}}
  & \cellcolor{green!20}\texttt{safe\_mode=True} &
    \cellcolor{green!20} JSON & 
    \cellcolor{green!20} - & 
    \cellcolor{green!20}\supported & 
    \cellcolor{green!20}\supported &
    \cellcolor{green!20}\cmark &
    \cellcolor{green!20}Blocks untrusted \texttt{Lambda} deserialization; custom objects must be registered \\
\addlinespace
  & \texttt{safe\_mode=False} &
    JSON & pickle &
    \supported & \supported &
    \xmark &
    Can restore models with unrestricted code; arbitrary code execution is possible by design \\
\addlinespace
  & Legacy &
    HDF5 & pickle &
    \supported & \supported &
    \xmark &
    Legacy format; can restore unrestricted code; arbitrary code execution is possible by design \\
\addlinespace
  & Weights-only &
    HDF5 & - &
    \supported & \notsupported &
    \cmark &
    Model must be provided separately (imported, downloaded, or copy–pasted) \\
\midrule
% --- TENSORFLOW BLOCK ---
\multirow[c]{4}{*}[-0.6ex]{TensorFlow~\cite{abadi2016tensorflow, TFSavedModelGuide, TFCheckpointGuide}}
  & \multirow[c]{3}{*}[-0.6ex]{\texttt{SavedModel}} &
    \multirow[c]{3}{*}[-0.6ex]{} protobuf + & & 
    \multirow[c]{3}{*}[-0.6ex]{\supported} & 
    \multirow[c]{3}{*}[-0.6ex]{\supported} &
    \multirow[c]{3}{*}[-0.6ex]{\xmark} &
    \multirow[c]{3}{*}[-0.6ex]{Encodes computational graphs and weights; arbitrary code execution is possible by design} \\
    & & raw data + & - & & & & \\
    & & assets & & & & & \\
\addlinespace
  & Checkpoint & %(\texttt{tf.train.Checkpoint}) &
    raw data & - &
    \supported & \notsupported &
    \cmark &
    For training checkpoints, not for sharing; model must be known and defined\\
\midrule
% --- PYTORCH BLOCK ---
\multirow{2}{*}{PyTorch~\cite{PyTorchFoundation, PyTorchSaveLoadRun}}
  & \texttt{weights\_only=False} &
    - & pickle &
    \supported & \supported &
    \xmark &
    Can restore models with unrestricted code; arbitrary code execution is possible by design \\
\addlinespace
  & \texttt{weights\_only=True} &
    - & pickle &
    \supported & \notsupported &
    \cmark &
    Restricted unpickler; model must be provided separately (imported, downloaded, or copy–pasted) \\
\midrule
% --- SCIKIT-LEARN BLOCK ---
\multirow{5}{*}{scikit-learn~\cite{pedregosa2011scikit, ScikitLearnModelPersistence}}
  & \multirow[c]{3}{*}[-0.6ex]{Pickle-based} &
    \multirow[c]{3}{*}[-0.6ex]{-} & pickle &
    \multirow[c]{3}{*}[-0.6ex]{\supported} & \multirow[c]{3}{*}[-0.6ex]{\supported} &
    \multirow[c]{3}{*}[-0.6ex]{\xmark} &
    \multirow[c]{3}{*}[-0.6ex]{Can restore models with unrestricted code; arbitrary code execution is possible by design} \\
    & & & joblib & & & & \\    
    & & & cloudpickle & & & & \\
\addlinespace
  & \cellcolor{green!20}With Skops &
    \cellcolor{green!20}JSON & 
    \cellcolor{green!20}- & 
    \cellcolor{green!20}\supported & 
    \cellcolor{green!20}\supported &
    \cellcolor{green!20}\cmark &
    \cellcolor{green!20}Allow-listed trusted types; flagged types require user review at load \\
\addlinespace
  & With ONNX &
    protobuf & - & 
    \supported & \partiallysupported &
    \cmark &
    Restricts the set of operations a model can use to implement its inference function \\
\midrule
% --- XGBOOST BLOCK ---
\multirow{3}{*}{XGBoost~\cite{chen2015xgboost, XGBoostSaveLoad}}
  & \multirow{2}{*}{Model} &
    JSON & \multirow{2}{*}{-} &
    \multirow{2}{*}{\supported} & \multirow{2}{*}{\partiallysupported} &
    \multirow{2}{*}{\cmark} &
    \multirow{2}{*}{No executable code is saved; to resume training, the hyperparameters must be provided separately} \\
    & & UBJSON &  & & & & \\    
\addlinespace
  & Model + hyperparams &
    - & pickle &
    \supported & \supported &
    \xmark &
    Can restore models with unrestricted code; arbitrary code execution is possible by design \\
\bottomrule
\end{tabular}
}
{\raggedright\scriptsize
\textsuperscript{$\star$}\, The \textit{Model} column is marked when the sharing format allows, in principle, to restore the complete model without manually redefining or reinstantiating it. Even when marked, some models may still require manual code definition (e.g., when they contain custom or non-standard objects).  

\noindent Legend: \supported\ = Supported, \partiallysupported\ = Partially supported (see \textit{Comments} column), \notsupported\ = Not supported, SO = Security-Oriented, \legendcmark = Yes, \legendxmark = No, \colorbox{green!20}{Green} = mechanisms that present security-oriented features while being self-contained. \par}

\vspace{0.4em}
\end{table*}

\section{Sharing Machine Learning Models}\label{sec:sharing}
% Machine learning (ML) model sharing can be examined from multiple perspectives, depending on how models are saved, distributed, and loaded. In this work, we study the security of ML model sharing. To scope our analysis, we first identify the most widely adopted ML frameworks (Section~\ref{sec:framework_level_sharing}) and model-sharing hubs (Section~\ref{sec:hub_level_sharing}). Using the official documentation of these frameworks, we extracted the recommended sharing formats and approaches, which we describe in detail below.

% Although TensorFlow~\cite{abadi2016tensorflow} is not strictly Python-based, the other selected frameworks are, and this aligns with Python’s dominance in ML and data science. Indeed, the 2024 JetBrains Developers Survey~\cite{jetbrains-python-survey-2024} reports that 34\% of Python developers primarily use it for ML tasks, while the 2022 Kaggle ML \& DS Survey~\cite{kaggle-survey-2022} found that 88.3\% of practitioners identify Python as their main programming language—far surpassing alternatives.

% To provide a comprehensive view of the model-sharing pipeline, we consider two sharing levels: the \textit{framework level}, which examines how frameworks handle model saving and loading; and the \textit{hub level}, which focuses on model-sharing platforms.

\ac{ML} model sharing can be examined from multiple perspectives, depending on how models are stored, distributed, and loaded. This work analyzes the security aspects of model sharing at two levels: the \textit{framework level}, which concerns how frameworks handle model serialization and loading, and the \textit{hub level}, which focuses on the distribution practices adopted by model-sharing platforms (\textit{hubs}). We focus on the most widely adopted \ac{ML} frameworks (Section~\ref{sec:framework_level_sharing}) and hubs (Section~\ref{sec:hub_level_sharing}), identifying their recommended sharing formats and practices from official documentation. Consequently, we do not consider third-party sharing mechanisms that are not officially endorsed by the frameworks and hubs analyzed in this study. This section provides the necessary background, while the security implications of model sharing are discussed in Section~\ref{sec:securityimplications}.

\subsection{Framework-level Sharing}\label{sec:framework_level_sharing}

% We conceptually classify framework-level model sharing techniques into two categories: \textit{self-contained} and \textit{non-self-contained} formats. This distinction captures whether the shared artifact (i.e., the model sharing files) includes all components required to fully restore and use a model. With a self-contained format, a user only needs to import the relevant framework and invoke its loading API to obtain a functional model. In contrast, non-self-contained formats typically store only configurations and weights. Therefore, users must define or import the corresponding model classes and instantiate the model object before loading parameters. A second, orthogonal classification concerns the file format itself. Some frameworks rely on \emph{code-based formats}, such as \texttt{pickle} or \texttt{joblib}, which directly serialize Python objects. Others employ \emph{data-based formats}, such as JSON, which declaratively describe model structure and allow reconstruction without executing embedded code.
% We classify framework-level model sharing along two dimensions. The first distinguishes between \textit{self-contained} formats, which include all components needed to restore a model via a single loading API, and \textit{non-self-contained} formats, which store only weights or configurations and require separate model definitions. The second distinguishes \emph{code-based formats} (e.g., pickle, joblib), which serialize Python objects directly, from \emph{data-based formats} (e.g., JSON), which declaratively encode information without executing any code.

During our preliminary analysis, we observed that framework-level model sharing formats can differ both in the content they store and in the way they represent a model. With respect to the stored content, we observed that some formats include all components needed to restore a model via a single loading API, while others store only weights or configurations and require a separate model definition. We refer to the former as \textit{self-contained} formats and to the latter as \textit{non-self-contained} formats. Regarding model representation, some formats serialize model code objects directly (e.g., pickle), whereas others describe model structures declaratively (e.g., JSON). We refer to the former as \textit{code-based} formats and to the latter as \textit{data-based} formats.

We then systematically analyze these two dimensions (i.e., stored content and model representation) across the five most widely adopted \ac{ML} frameworks, as identified in the 2022 Kaggle \ac{ML} \& DS Survey~\cite{kaggle-survey-2022}: TensorFlow~\cite{abadi2016tensorflow}, Keras~\cite{chollet2015keras}, PyTorch~\cite{PyTorchFoundation}, scikit-learn~\cite{pedregosa2011scikit}, and XGBoost~\cite{chen2015xgboost}. Notably, TensorFlow is the only one in this group that is not strictly Python-based, highlighting the dominance of Python in \ac{ML} fields. According to the same survey, 88.3\% of \ac{ML} practitioners identified Python as their primary programming language, far surpassing all alternatives.  This distribution is further confirmed by the 2024 JetBrains Python Developers Survey~\cite{jetbrains-python-survey-2024}, which reports usage rates of 48\% for TensorFlow, 30\% for Keras, 60\% for PyTorch, 67\% for scikit-learn, and 22\% for XGBoost.

We now analyze the model-sharing techniques suggested by the official documentation of each selected framework. Table~\ref{tab:framework-level-sharing} summarizes their main characteristics and security implications, discussed in Section~\ref{sec:securityimplications}.

\subsubsection{Keras}\label{sec:keras_sharing}

% Keras~\cite{chollet2015keras} is a high-level deep learning API. Originally developed as an independent project, it is now tightly integrated into TensorFlow~\cite{abadi2016tensorflow}, where it serves as the default high-level interface.

% Keras provides an API~\cite{KerasSavingLoading} for saving and loading models, enabling users to preserve architecture, weights, and training configuration, and to easily share pretrained models. Except for specific cases involving \texttt{Lambda} layers, a model can typically be restored without requiring any prior class or function definitions, thereby offering a truly self-contained solution.
Keras is a \acl{DL} API~\cite{chollet2015keras} tightly integrated into TensorFlow~\cite{abadi2016tensorflow}, where it serves as the default high-level interface. Keras provides APIs~\cite{KerasSavingLoading} for saving and loading models, enabling users to preserve architecture, weights, and training configuration.

\mypar{Self-contained Formats}
The recommended persistence format is the \texttt{.keras} archive, which has been introduced in Keras 2.11. It is a ZIP file whose core element is a \texttt{config.json}, which describes the model’s configuration and architecture in a hierarchical JSON structure. This data-based representation typically allows models to be restored without requiring prior class or function definitions. Keras provides a \texttt{safe\_mode} option (introduced in v2.13) to restrict insecure deserialization during model loading~\cite{KerasSavingLoading}. Its documentation states:
% \begin{list}{}{\setlength{\leftmargin}{10pt}%
%                \setlength{\rightmargin}{10pt}%
%                \setlength{\topsep}{6pt}%
%                \setlength{\parsep}{6pt}%
%                \setlength{\partopsep}{0pt}}
% \item
% \textit{``\texttt{safe\_mode}: Boolean, whether to disallow unsafe lambda deserialization. 
% When \texttt{safe\_mode=False}, loading an object has the potential to trigger arbitrary code execution. 
% This argument is only applicable to the Keras v3 model format. Defaults to \texttt{True}.''}~\cite{KerasSavingLoading}
% \end{list}

\smallskip\noindent
\textit{``\texttt{safe\_mode}: Boolean, whether to disallow unsafe lambda deserialization. 
When \texttt{safe\_mode=False}, loading an object has the potential to trigger arbitrary code execution. 
This argument is only applicable to the Keras v3 model format. Defaults to \texttt{True}.''}~\cite{KerasSavingLoading}
\smallskip

\noindent
This option addresses a key risk: Keras supports \texttt{Lambda} layers, which can wrap arbitrary Python expressions as \texttt{Layer} objects. When serialized, these layers may embed Python bytecode representing custom logic. If \texttt{safe\_mode} is disabled, this bytecode is deserialized and will execute at load time. In such cases, the \texttt{.keras} format becomes hybrid, combining data with executable code. Additionally, developers may allow custom, potentially \textit{``unsafe''} types by using a specific decorator~\cite{KerasSavingLoading}.
In earlier versions, complete models were saved in the HDF5 format (\texttt{.h5}). Although this format also stores architecture, weights, and training configuration in a single file, it is now advertised as \textit{``legacy''}: in Keras 3, loading an \texttt{.h5} file raises a warning.

\mypar{Non-self-contained formats} Keras supports weights-only persistence through the \texttt{save\_weights} and \texttt{load\_weights} methods~\cite{KerasWeightsSaving}. In this case, the weights are stored in a \texttt{.h5} file. However, restoring them requires instantiating the model architecture beforehand, meaning that the file alone is insufficient for full reconstruction.

% The recommended sharing format is the \texttt{.keras} ZIP archive, which is primarily data-based. The core of the model structure is stored in a \texttt{config.json} file inside the archive, which captures the model's configuration and architecture in a hierarchical (tree-like) JSON format. 

% Keras also includes a \texttt{safe\_mode} option intended to restrict insecure deserialization behavior during model loading~\cite{KerasSavingLoading}. However, when \texttt{safe\_mode} is disabled, the format also permits embedding Python bytecode—particularly within \texttt{Lambda} layers—using pickle. As a result, the format effectively becomes hybrid, combining declarative data-based structure with executable code-based elements.

\subsubsection{TensorFlow}

TensorFlow is an open-source \acl{DL} framework developed by Google~\cite{abadi2016tensorflow}. TensorFlow and Keras~\cite{chollet2015keras} complement each other: Keras serves as the high-level API for building and training models, while TensorFlow provides the low-level execution engine and additional abstractions. This relationship extends to model persistence. For models built with Keras layers, the recommended format is the \texttt{.keras} archive (see Section~\ref{sec:keras_sharing})~\cite{TFKerasSaveLoad}. By contrast, TensorFlow’s native format, \texttt{SavedModel}, is designed for interoperability and deployment scenarios, particularly when working with raw TensorFlow objects or non-Python environments~\cite{TFSavedModelGuide}.
Excluding Keras, which has already been discussed, TensorFlow supports two persistence mechanisms: \texttt{SavedModel} and \textit{training checkpoints}~\cite{TFKerasSaveLoad,TFSavedModelGuide}.

\mypar{Self-contained formats} The \texttt{SavedModel} format stores models in a directory containing a protocol buffer encoding a computation graph,  weights (in the training checkpoints format), and optional files. Models can be restored without requiring the original model object definition.

\mypar{Non-self-contained formats}  
% TensorFlow supports training checkpoints, which record only the weight values. As the official guide states, checkpoints require the original model to be defined before restoration~\cite{TFCheckpointGuide}.
Training checkpoints store only the model’s weight values. As stated in the official guide, restoring a checkpoint requires the original model to be defined beforehand~\cite{TFCheckpointGuide}.

\subsubsection{PyTorch}

PyTorch is an open-source \acl{DL} framework developed by the PyTorch Foundation~\cite{PyTorchFoundation}. It offers two main persistence options~\cite{PyTorchSaveLoadRun,PyTorchSerialization}: loading the model object configuration or loading only its state dictionary (i.e., the parameters learned during training). This behavior is controlled via the \texttt{weights\_only} flag. In both cases, serialization is performed using Python’s pickle module (code-based format).

\mypar{Self-contained formats} When \texttt{weights\_only=False} and the model is not defined as a custom class, PyTorch relies on raw pickle serialization, allowing the model to be loaded without a prior class definition in the environment. However, in this case, no security restrictions are enforced to prevent arbitrary code execution, as raw pickle is known to be insecure~\cite{PythonPickleDocs}.

\mypar{Non-self-contained formats} \texttt{weights\_only=True} requires the instantiation of a model on which to load the weights. Additionally, if the model is defined using a custom class, even when \texttt{weights\_only=False}, its definition must still be available at load time, although instantiation is handled internally. Interestingly, when \texttt{weights\_only=True}, PyTorch uses pickle with a restricted unpickler that limits deserialization to \texttt{torch.Tensor} objects and primitive types, and prevents dynamic imports during loading.

\subsubsection{scikit-learn}\label{sec:scikit}

scikit-learn is an open-source Python library for \textit{traditional} \ac{ML}~\cite{pedregosa2011scikit}. The official documentation~\cite{ScikitLearnModelPersistence} describes several persistence approaches: (i) Python’s pickle, joblib, cloudpickle modules, (ii) the Skops library~\cite{JalaliBossanMerveSkops}, and (iii) conversion to the framework-agnostic ONNX format~\cite{onnx}. All of these are self-contained approaches, and we describe each of them in the following.

\mypar{Pickle/Joblib/Cloudpickle} The traditional scikit-learn model persistence approach uses pickle, joblib, or cloudpickle. These formats serialize estimators in a code-based manner: unrestricted deserialization executes Python bytecode, posing risks of arbitrary code execution~\cite{ScikitLearnModelPersistence}.

% \mypar{Skops} To mitigate these risks, Hugging Face introduced Skops in 2022~\cite{JalaliBossanMerveSkops}, with native support for publishing on the Hugging Face Hub~\cite{HuggingFaceHub}. Skops avoids executing arbitrary code during loading and enforces a trusted type validation step~\cite{SkopsPersistence}. Its \texttt{.skops} archive contains a JSON schema (\texttt{schema.json}) describing the estimator’s structure and parameters, making it a self-contained, data-based format, similar in design to Keras’s \texttt{.keras} archive. 

\mypar{Skops} 
Hugging Face introduced Skops in 2022~\cite{JalaliBossanMerveSkops}, with native support for publishing models on their hub~\cite{HuggingFaceHub}. According to the official documentation~\cite{SkopsPersistence}, Skops provides \texttt{skops.io.dump()} and \texttt{skops.io.load()} as secure alternatives, designed to prevent arbitrary code execution and to reject unknown or malicious objects by default. The \texttt{.skops} archive is a ZIP file that includes, among other elements, a JSON schema (\texttt{schema.json}) describing the estimator’s structure in a tree-like format, with nodes such as \texttt{MethodNode}, \texttt{TypeNode}, or \texttt{FunctionNode}. This design makes the \texttt{.skops} a data-based format, similar in spirit to Keras’s \texttt{.keras} archive. The intended workflow involves calling \texttt{get\_untrusted\_types()} to inspect the model and identify objects that are not trusted by default. Users must then manually review these objects and explicitly allowlist them when calling \texttt{load()}; otherwise, loading fails. Therefore, security in Skops depends not only on its data-based design, but also on users' reviewing capabilities. %While it avoids executing arbitrary code during deserialization, the library remains under active development and may still be subject to unresolved security issues.

\mypar{ONNX} scikit-learn models can be exported to the Open Neural Network Exchange (ONNX)~\cite{onnx} format using the \texttt{skl2onnx} converter~\cite{ScikitLearnModelPersistence}. ONNX defines a framework-agnostic format based on protocol buffers and provides a purely data-based representation, making models self-contained and portable across languages and environments. Unlike other self-contained approaches, the original estimator object (with its class structure and custom code) is not preserved. Instead, the model is reduced to a computational graph and its parameters, where the graph can only contain a limited set of predefined \ac{ML} operators. ONNX serialization depends on the coverage of the chosen conversion tool, and many scikit-learn estimators remain unsupported.

\subsubsection{XGBoost}

XGBoost~\cite{chen2015xgboost} is a widely used gradient boosting framework. It provides dedicated APIs for saving and loading models in either JSON or UBJSON (Universal Binary JSON) formats~\cite{XGBoostPythonAPI}.

\mypar{Self-contained formats} Unlike \acl{DL} frameworks, XGBoost implements gradient boosting, which has a \textit{closed} structure. As a result, importing the standard file format is sufficient to treat it as self-contained. However, in scenarios such as distributed or collaborative learning, where training-critical hyperparameters are not included in the JSON file (e.g., the \texttt{max\_depth} parameter), the framework’s authors recommend using the built-in pickle format to fully serialize the \texttt{Booster} object~\cite{XGBoostSaveLoad}. Alternatively, the hyperparameters must be set and provided separately.

\begin{table*}[th]
\centering
\small
\setlength{\tabcolsep}{5pt}
\renewcommand{\arraystretch}{1.2}

\caption{
% \marco{"Green highlighting not clearly defined (same issue with Table 2)". Table caption could be more expressive (even more important for Table 2).}Hub-level sharing: documented characteristics and security considerations.
Hub-level ML model sharing mechanisms and security considerations, summarizing the characteristics of the analyzed model-sharing hubs and indicating their hosting model, reference format (i.e., the specific model format the hub relies on, if any), and their associated security models.
}
\label{tab:hub-level-sharing}
\resizebox{\linewidth}{!}{%
\begin{tabular}{@{}%
l  % Hub
c c  % Characteristics flags
c  % SO
l  % Comments
@{}}
\toprule
\multirow{2}{*}{\textbf{Hub}} & \multicolumn{2}{c}{\textbf{Characteristics}} & \multicolumn{2}{l}{\textbf{Security model (security posture + commentary)}} \\
\cmidrule(lr){2-3}\cmidrule(lr){4-5}
& \textbf{Centralized hosting} & \textbf{Reference format} & \textbf{SO} & \textbf{Comments} \\
\midrule
\cellcolor{green!20}Hugging Face Hub~\cite{HuggingFaceHub} &  \cellcolor{green!20}\supported &  \cellcolor{green!20}- &  \cellcolor{green!20}\cmark &  \cellcolor{green!20}Malware, pickle, and secret scanning; integrates third-party model scanners\\
\midrule
Kaggle Models~\cite{KaggleModels} & \supported & - & \xmark & No explicit security measures documented; relies on notebook isolation and standard formats \\
\midrule
PyTorch Hub~\cite{PyTorchHub} & \notsupported & - & \xmark & No explicit security measures documented; arbitrary Python execution from repositories \\
\midrule
TensorFlow Hub~\cite{TensorFlowHub} & \supported & \texttt{SavedModel} & \xmark & No explicit security measures documented \\
\midrule
Keras Hub~\cite{kerashub2024} & \supported & \texttt{.keras} & \xmark & No explicit security measures documented  \\
\bottomrule
\end{tabular}
}

\vspace{0.4em}
{\raggedright\scriptsize
Legend: SO = Security-Oriented, \supported\ = Supported, \notsupported\ = Not supported, \legendcmark = Yes, \legendxmark = No. \colorbox{green!20}{Green} = mechanisms that present security-oriented features.\par
}
\end{table*}

\subsection{Hub-level Sharing}\label{sec:hub_level_sharing}

% Model-sharing platforms (or hubs) provide centralized infrastructures where users can publish, discover, and download pre-trained models. While these platforms accelerate research and application development, they also introduce unique security considerations deriving from their hosting policies, model verification, and distribution mechanisms.

An additional layer of the ecosystem is represented by model-sharing hubs. These provide infrastructures where users can publish and download pre-trained models, accelerating research and application development. Moreover, they may also incorporate further safeguards—such as content scanning, model verification, or curation policies—to protect users against threats beyond those arising from local framework-level loading.
As with the frameworks, for our analysis we selected these widely used hubs: Hugging Face Hub~\cite{HuggingFaceHub}, Kaggle Models~\cite{KaggleModels}, TensorFlow Hub~\cite{TensorFlowHub}, Keras Hub~\cite{kerashub2024}, and PyTorch Hub~\cite{PyTorchHub}. This choice is supported by usage statistics: the 2022 Kaggle \ac{ML} \& DS Survey~\cite{kaggle-survey-2022} ranks these among the most popular \ac{ML} model repositories. Table~\ref{tab:hub-level-sharing} summarizes their characteristics along with their security implications (discussed in Section~\ref{sec:securityimplications}).

\subsubsection{Hugging Face Hub} Hugging Face Hub~\cite{HuggingFaceHub} is one of the most popular open hubs for sharing \ac{ML} models. It hosts over 2.1 million models in a Git-based repository format, supporting a variety of file types. The platform implements multiple security measures~\cite{HuggingFaceHubSecurity}: every uploaded file is automatically scanned for malware using ClamAV~\cite{ClamAV}, and any pickle-based model file is ``pickle-scanned'' to enumerate suspicious included modules. The Hub also performs secret scanning to detect accidentally leaked credentials. Furthermore, Hugging Face integrates two third-party model scanning services: Protect AI~\cite{ProtectAI} and JFrog~\cite{JFrog}.

% \subsubsection{Kaggle Models} The Kaggle repository~\cite{KaggleModels} supports both public and private model uploads from users and companies, often complemented by configuration files or metrics, and is tightly integrated with Kaggle Notebooks. 
\subsubsection{Kaggle Models} The Kaggle repository~\cite{KaggleModels} supports both public and private model uploads, often complemented by configuration files or metrics, and is tightly integrated with Kaggle Notebooks. 
Many models, including those from TensorFlow Hub~\cite{TensorFlowHub} (which fully migrated to Kaggle Models in 2023), can be loaded through high-level APIs. Security relies primarily on the isolation of Kaggle’s cloud notebook environment and the use of standard model formats, as the platform does not publicly document malware or pickle scanning for user-contributed models.

\subsubsection{PyTorch Hub} PyTorch Hub~\cite{PyTorchHub} is a built-in model-sharing mechanism for PyTorch that enables users to load models from GitHub repositories with a single command~\cite{PyTorchHub}. Unlike centralized platforms, it is entirely decentralized: models are hosted on GitHub, and PyTorch Hub collects only an entry-point script (\texttt{hubconf.py}) for fetching and instantiating them. The PyTorch team maintains a list of models on the official PyTorch Hub page (via a pull-request submission process)~\cite{PyTorchHubDocs}, but using the \texttt{torch.hub} APIs, it is possible to load models from any public repository by URL. This flexibility means that \texttt{torch.hub.load()} downloads the target repository and executes its model-loading code. PyTorch Hub performs no automated security scanning or integrity checks.

\subsubsection{TensorFlow Hub} 
TensorFlow Hub~\cite{TensorFlowHub} was originally released alone as a central repository for reusable TensorFlow modules. It provides a library and hosting platform where models can be published and imported via simple APIs, such as  \texttt{hub.KerasLayer}. Unlike decentralized solutions, TensorFlow Hub offers a collection of models maintained either by Google or by community contributors. Since 2023, TensorFlow Hub has been fully integrated into Kaggle Models~\cite{KaggleModels}, and while the API remains available, models are uploaded directly to Kaggle. By default, it adopts the \texttt{SavedModel} format without additional restrictions and makes no public mention of automated scanning.

% Security measures are not explicitly documented: the platform primarily relies on the safety guarantees of the \texttt{SavedModel} format and on the controlled hosting infrastructure, without public reference to automated malware or pickle scanning.

\subsubsection{Keras Hub} 
Keras Hub~\cite{kerashub2024}, introduced with Keras 3, provides a collection of ready-to-use pretrained Keras models. Unlike TensorFlow Hub, which hosts TensorFlow-specific modules, Keras Hub focuses exclusively on models designed to integrate seamlessly with the Keras API. Models can be loaded directly via \texttt{.from\_preset()} and are distributed in the standardized format \texttt{.keras}. No content scanning is explicitly documented.

\section{Model Sharing Security Implications (RQ1)}
\label{sec:securityimplications}

% \sk{sia qui che dopo scrivete "in section 2 (la precedente) we identify... mantenendo il presente. Siete sicuri? secondo me dovrebbe essere al passato, o comunque io lo ho sempre usato}
% \jh{io sempre il presente, tranne nelle conclusioni :D}
%In Section~\ref{sec:sharing}, we identify several frameworks and sharing hubs based on their popularity and analyze their respective sharing approaches. In this section, we analyze the security implications of these approaches, with the goal of clarifying the current state of security in \acf{ML} model sharing and highlighting critical issues that deserve further investigation. Before conducting the security analysis, we formally define the threat model underlying our study.

This section analyzes the security implications of the sharing approaches adopted by frameworks and sharing hubs described in Section~\ref{sec:sharing}, aiming to clarify the current state of security in \ac{ML} model sharing and highlight critical issues requiring further investigation. Before the analysis, we formally define the threat model underlying our study.

\subsection{Threat Model}\label{sec:threat_model}

We consider the threat posed by malicious \ac{ML} model artifacts that target users loading models through popular \ac{ML} frameworks. Our threat model defines the attacker’s target, objectives, and capabilities. Finally, we discuss the relevance of our threat model in the real world.

\mypar{Attacker's Target} Our analysis focuses on the \ac{ML} model-loading pipeline executed locally on a user’s machine. The system the attacker targets consists of: (i) the \emph{user environment}, including the operating system with an installed \ac{ML} framework (e.g., PyTorch~\cite{PyTorchFoundation}, TensorFlow~\cite{abadi2016tensorflow}, Keras~\cite{chollet2015keras}, scikit-learn~\cite{ScikitLearnModelPersistence}); (ii) the \emph{model artifact}, a serialized pre-trained file such as \texttt{.pth}, \texttt{.keras}, \texttt{.h5}, or \texttt{.pkl}; and (iii) the \emph{loading mechanism}, i.e., framework-provided functions such as \texttt{torch.load()}~\cite{PyTorchSaveLoadRun}. We focus on scenarios where a user loads a model obtained from an external source.

\mypar{Attacker's Goal} We consider an attacker who crafts a malicious model artifact to compromise the victim’s system upon loading. The primary goal is to achieve arbitrary code execution. Such attacks typically exploit vulnerabilities in the deserialization process, targeting the host system itself rather than the functionality of the model.

\mypar{Attacker's Capabilities} The attacker can create, modify, and distribute \ac{ML} model artifacts but has no prior access to the target’s system and cannot influence the victim’s environment, configuration settings, or model-loading flags.
We consider two main distribution channels: (1) \emph{public repository poisoning}, where a malicious model is uploaded to a trusted platform (e.g., Hugging Face Hub~\cite{HuggingFaceHub}, Kaggle, GitHub) under the mask of a legitimate resource; and (2) \emph{direct delivery}, where the artifact is sent to the victim via private channels such as email or cloud storage. 
% Formally, we model the loading function as:
% \[\text{Load}: m \rightarrow (M, S)\]
% For a benign artifact $m$, $\text{Load}(m)$ returns a model $M$ and preserves the system state $S$. For a malicious artifact $m^*$, $\text{Load}(m^*)$ returns a (possibly intended) model $M$ but produces a compromised state $S'$, which encloses unauthorized system access. This formalization emphasizes that the primary target is the host system rather than the model’s functionality.

\mypar{Real-World Relevance} Our threat model is grounded in recent large-scale measurements showing that malicious model artifacts exist in practice (e.g., 91 malicious models and several poisoned dataset scripts discovered by monitoring Hugging Face over a three-month period)~\cite{zhao2024models, casey2024large}. While these studies focus only on known vulnerabilities in model deserialization (e.g., Python’s pickle unsafe deserialization~\cite{PythonPickleDocs}), they demonstrate the real-world relevance of our threat model. Furthermore, according to a recent Hugging Face report, Protect AI~\cite{ProtectAI} has scanned over 4.47 million unique model versions across 1.41 million repositories, identifying 352{,}000 unsafe or suspicious issues in 51{,}700 models as of April~2025~\cite{MorganPAI6Months}.

\subsection{Security Analysis}

The frameworks described in Section~\ref{sec:sharing} adopt or recommend different strategies for model sharing. On top of these mechanisms, model hubs provide their own approaches for distributing pre-trained models.
Here, we systematically analyze whether and how frameworks and hubs address the risks defined in the threat model, starting from the official documentation they provide. In particular, we classify \textit{framework-level} and \textit{hub-level} mechanisms as \textit{security-oriented} (directly claiming or implicitly providing security properties) or \textit{non-security-oriented}. Tables~\ref{tab:framework-level-sharing} and~\ref{tab:hub-level-sharing} summarize the classification and corresponding security models.

\secanf{Framework-level Security Analysis}
\label{saf}

\mypar{Security-Oriented Formats}
Keras and scikit-learn provide formats embedding the entire model object (self-contained approach) that are explicitly presented as security-oriented in their documentation: \texttt{safe\_mode=True} for Keras~\cite{KerasSavingLoading} and Skops for scikit-learn~\cite{ScikitLearnModelPersistence}. Keras's \texttt{safe\_mode} flag disables \textit{``unsafe lambda deserialization''}~\cite{KerasSavingLoading}, thereby preventing execution of pickle-encoded payloads. Skops enforces a trusted type validation step~\cite{SkopsPersistence}, requiring users to manually review and approve potentially insecure objects before loading (via \texttt{get\_untrusted\_types()}). The developers describe this mechanism as \textit{``secure persistence''}~\cite{SkopsPersistence}, but also explicitly caution that the library is still under active development and may contain unresolved security issues. Both Keras and Skops base their security models on avoiding pickle in favor of declarative, JSON-based formats. 
However, our analysis in Section~\ref{sec:whensecisnotsec} shows that this assumption does not always hold.
In particular, although JSON itself is a data-based format, the way Keras and Skops process their JSON-based files effectively makes them behave like code-based formats—creating critical issues discussed later in the paper.

% While this design choice intuitively suggests stronger security and shapes users’ perception of security (Section~\ref{sec:user-perception}), our analysis in Section~\ref{sec:whensecisnotsec} shows that this assumption does not always hold. In particular, although JSON itself is a data-based format, the way Keras and Skops process their JSON-based files effectively makes them behave like code-based formats—creating critical issues discussed later in the paper.

XGBoost uses a JSON/UBJSON-based format~\cite{XGBoostSaveLoad} that saves no executable code. To resume training, hyperparameters must be retrieved and set separately, while the architecture code is standardized and provided by the XGBoost library itself. Security is implicitly provided by the fixed architecture of the XGBoost model. Notably, the developers do not advertise this design as security-oriented. Furthermore, the documentation explicitly recommends raw pickle serialization in distributed or collaborative scenarios (where preserving hyperparameters is needed to resume training), reintroducing well-known risks of arbitrary code execution.

ONNX~\cite{onnx} does not explicitly make security claims in its documentation. However, scikit-learn refers to ONNX as the \textit{``most secure solution''} for model persistence~\cite{ScikitLearnModelPersistence}. This characterization derives from ONNX’s operator-based design (see Section~\ref{sec:framework_level_sharing}), which inherently limits the attack surface. Its restricted expressiveness—a limited set of predefined operators, optionally extendable through external libraries—makes arbitrary code execution unlikely. Nevertheless, while ONNX is well-suited for inference (scikit-learn recommends ONNX only for that purpose~\cite{ScikitLearnModelPersistence}), it is far less flexible for training.

Non-self-contained formats such as PyTorch's \texttt{weights\_only} serialization, Keras's weights-only API, and TensorFlow’s training checkpoints can also be considered relatively security-oriented. 
% Note that PyTorch recommends the weights-only mode to reduce the risk of arbitrary code execution, thereby claiming security guarantees. 
Note that PyTorch explicitly recommends the \texttt{weights\_only} mode in its documentation, claiming the presence of security measures that are not enforced when this mode is disabled.
These approaches store only numerical parameters, making them inherently security-oriented. However, they merely shift the trust problem: model architecture code must be provided separately, and if obtained from unverified sources, it reintroduces the same arbitrary code execution risks.

\mypar{Non-Security-Oriented Formats}
Other frameworks rely on inherently insecure formats. Keras \texttt{safe\_mode=False} or legacy HDF5, PyTorch \texttt{weights\_only=False}, scikit-learn (pickle-based), and XGBoost (model + hyperparameters) all persist models using unrestricted pickles. %, exposing users to arbitrary code execution. 
TensorFlow’s \texttt{SavedModel} format~\cite{TFSavedModelGuide}, despite being based on computation graphs, is explicitly described by TensorFlow developers as insecure when loading untrusted models~\cite{TFSecurityPolicy}.
% \begin{list}{}{\setlength{\leftmargin}{10pt}%
%                \setlength{\rightmargin}{10pt}%
%                \setlength{\topsep}{10pt}%
%                \setlength{\parsep}{10pt}%
%                \setlength{\partopsep}{10pt}}
% \item\textit{"When loading untrusted serialized computation graphs (in the form of a \texttt{GraphDef}, \texttt{SavedModel}, or equivalent on-disk format), the set of computation primitives available to TensorFlow is powerful enough that you should assume that the TensorFlow process effectively executes arbitrary code."}~\cite{TFSecurityPolicy}
% \end{list}

% \smallskip\noindent
% \textit{"When loading untrusted serialized computation graphs (in the form of a \texttt{GraphDef}, \texttt{SavedModel}, or equivalent on-disk format), the set of computation primitives available to TensorFlow is powerful enough that you should assume that the TensorFlow process effectively executes arbitrary code."}~\cite{TFSecurityPolicy}
% \smallskip

\secanh{Hub-level Security Analysis}
\label{sah}

\mypar{Security-Oriented Hubs}
Among the analyzed hubs, Hugging Face Hub~\cite{HuggingFaceHub} stands out as the only platform with active and documented security measures. These include malware scanning, pickle scanning, and secret scanning, complemented by integrations with external services such as Protect AI~\cite{ProtectAI} and JFrog~\cite{JFrog}. When a model artifact is uploaded, it is subjected to these scans, and if no issues are detected, the platform marks the model with a ``Safe'' label. Overall, this approach reflects Hugging Face’s recognition of models as executable code and its commitment to enforcing security practices consistent with that perspective.

\mypar{Non-Security-Oriented Hubs} 
TensorFlow Hub~\cite{TensorFlowHub} and Keras Hub~\cite{kerashub2024} do not perform systematic artifact scanning; instead, they offload security to the framework level by relying on controlled formats (\texttt{.keras}, \texttt{SavedModel}). Kaggle benefits from community curation and the isolation of its notebook environment, though unverified models can still be uploaded. 
PyTorch Hub provides no hub-level protections at all: it executes arbitrary Python code from repositories, leaving responsibility entirely to the user.

\section{When \textit{secure} is not secure (RQ2)}
\label{sec:whensecisnotsec}

While existing evidence, both from prior scientific studies~\cite{zhao2024models, casey2024large} and from independent checks by model hubs~\cite{MorganPAI6Months}, has brought attention to the relevance and scale of attacks targeting \ac{ML} model loading mechanisms, these efforts have largely focused on well-known attack vectors.
In particular, they have examined inherently insecure methods, such as pickle deserialization, but have not questioned the effectiveness of mechanisms that are claimed to be secure.
In this section, we go a step further: building on the observations from our security analysis in Section~\ref{sec:securityimplications}, we evaluate how closely the security narrative promoted by popular \ac{ML} frameworks and hubs, claiming to secure the entire model-sharing process, matches reality.

As discussed, some approaches are deliberately \textit{non-security-oriented}, relying on insecure formats or distribution methods. Others offer limited security but at the expense of flexibility or by shifting responsibility to users or other workflow components (e.g., requiring the model architecture to be separately obtained). 
A smaller subset explicitly presents itself as \textit{security-oriented}, seeking to make model sharing more secure overall. 
This section focuses on hubs that provide scanning mechanisms and on frameworks that claim security combined with complete model object sharing. In particular, from a framework-level perspective, we assess the solutions highlighted in Table~\ref{tab:framework-level-sharing}: the \texttt{safe\_mode} in Keras and the \textit{``secure persistence''}~\cite{SkopsPersistence} of Skops for scikit-learn. From a hub-level perspective, we analyze the only hub that provides embedded scanning mechanisms, as highlighted in Table~\ref{tab:hub-level-sharing}: the Hugging Face Hub.

Notably, our analysis uncovered six 0-day vulnerabilities (all assigned to CVEs) across Keras and Skops, each allowing arbitrary code execution during model loading. % (i.e., before the model invocation). 
Interestingly, both frameworks base their security claims on data-based format persistence, which our findings show does not hold in practice. 
Moreover, our hub-level experiments demonstrated that it is feasible for an attacker to distribute exploits leveraging those vulnerabilities without being detected by the scanning tools integrated into Hugging Face.

\mypar{Vulnerability Research Methodology} 
% All the vulnerabilities presented in this work were discovered through manual reverse engineering and code analysis, focusing specifically on the parts of the Keras and Skops open-source codebases related to the security mechanisms we aimed to assess. We analyzed the latest available versions on both PyPI and GitHub at the time of our study. As detailed in the following sections, the vulnerabilities arise from logical flaws and missing (or insufficiently strict) validation checks, ultimately allowing attackers to bypass the expected security guarantees.
All vulnerabilities described below were discovered through manual reverse engineering of the latest Keras and Skops open-source codebases from GitHub, focusing on security-related functionalities.

% This approach is consistent with the goal of the paper, implementing a hypothesis-driven adversarial methodology in which each documented security claim is treated as a hypothesis and tested under the threat model defined in Section~\ref{sec:threat_model}. In this sense, each discovered vulnerability serves as a counterexample to the corresponding security claim.

% As detailed in the following sections, the vulnerabilities result from logical flaws and missing or insufficiently strict validation checks, allowing attackers to bypass intended security guarantees.

\mypar{Pre-existing CVEs}
% {\color{red}Before presenting our findings, we briefly review prior public CVEs for Keras and skops as an indicator of their security maturity and the historical attention given to model persistence risks. As regards Keras, no CVEs had been assigned to the \texttt{.keras} format or \texttt{safe\_mode} prior to our work. For skops, we found only one. This suggests that the security of these secure model-sharing mechanisms has received limited attention and has not been publicly evaluated in depth. A full discussion is provided in Appendix~\ref{appendix:cve}.}
We reviewed publicly available CVEs for Keras and Skops. Prior to our work, no CVEs had been assigned to the \texttt{.keras} format or \texttt{safe\_mode}, and only one to Skops. A detailed analysis is provided in Appendix~\ref{appendix:cve}.
%{\color{orange}This suggests a historical lack of attention to model persistence mechanisms that are claimed to be secure.}
%This suggests that the security of secure model-sharing mechanisms have received limited attention.

\subsection{Keras}
% We analyze the \texttt{safe\_mode} of Keras to evaluate how effectively it prevents arbitrary code execution in practice.
% We discovered three distinct vulnerabilities enabling arbitrary code execution, detailed in the following subsections. None of these attacks requires the victim to define custom objects—no user-defined classes, functions, or decorators are needed. 

% Our findings demonstrate that an attacker can craft a malicious model file which, when simply loaded by Keras, triggers code execution \emph{before} the model is ever used.
%In what follows, we refer to each vulnerability as \texttt{KV.\#}, where \texttt{KV} denotes "Keras Vulnerability" and \texttt{\#} is a sequential identifier.

\vksec{Abusing Insecure Module Resolution}
\label{keras_first}
We discovered that due to insufficient validation in the Keras model loading process, a carefully crafted \texttt{config.json} file inside a \texttt{.keras} model archive can specify insecure Python modules and functions to be imported and executed during loading.
In particular, an attacker can build a \texttt{config.json} such that, for example, \texttt{subprocess.run} is interpreted as a model layer. Keras’s loading logic performs minimal validation: if the specified class name resolves to a Python \texttt{FunctionType}—which \texttt{subprocess.run} does—no further checks are performed.
Arguments can then be passed abusing the input–output relations within Keras’s internal computation graph, enabling execution of commands such as \texttt{subprocess.run("/bin/sh")}. Since the model layer is instantiated at load time, the command is executed during loading. A simplified \texttt{config.json} snippet is provided in the appendices (Listing~\ref{fig:malicious-config}), while the complete \ac{PoC} is available on GitHub. Crucially, this finding demonstrates the existence of entirely different exploitation paths in Keras's loading mechanism, beyond the abuse of \texttt{Lambda} layers.

\mypar{Disclosure}
After coordinated disclosure, the vulnerability was assigned the identifier \texttt{CVE-2025-1550} (CVSS 7.3, CNA: Google LLC)~\cite{CVE-2025-1550}. To the best of our knowledge, this is the first CVE assigned to Keras’s model loading mechanism after the introduction of \texttt{safe\_mode}, and thus the first to demonstrate a weakness in this security feature. The issue was mitigated in Keras version 3.9 through the introduction of stricter validation, which restricts imports to a limited set of trusted modules, primarily within Keras itself (allowlist approach).

% Following the CVE, ProtectAI’s artifact scanner—also integrated into Hugging Face—was publicly updated with a new threat definition specifically targeting this finding~\cite{MorganPAI6Months-anon,ProtectAI-KERAS-301-anon}.

% A mitigation introducing stricter validation checks on the \texttt{module} key in \texttt{config.json} was included in Keras version 3.9.

\vksec{Code Reuse via \texttt{Lambda} Layers}
\label{keras_second}
% \begin{marcosbox}
% \linkedtakeaways{TK.1 – Misplaced Trust in Format-Based Security; TK.1 – Misplaced Trust in Format-Based Security}
% \end{marcosbox}
% \texttt{Lambda} layers in Keras can be used not only to serialize arbitrary Python bytecode but also to reference functions from specified Python modules. 
% As a result, we found that the restriction introduced following the disclosure of \texttt{CVE-202X-XXX1}, as described above, does not fully prevent arbitrary code execution.

% In particular, we demonstrate how an attacker can achieve arbitrary code execution through \emph{code reuse} by leveraging \texttt{Lambda} layers that invoke legitimate functions from Keras’s internal modules. As a \ac{PoC}, we crafted a model that disables \texttt{safe\_mode} during loading, even if it was initially enabled by the user, by invoking an internal Keras utility. Listing~\ref{fig:disable-safe-mode} shows a snippet of the corresponding \texttt{config.json}, where arguments are passed both through the top-level \texttt{inbound\_nodes} key and the \texttt{arguments} key within the configuration of the \texttt{Lambda} layer. Then, by relying on other internals, such as the function used to load the model (now with \texttt{safe\_mode} disabled), we show how to achieve full code execution. This represents just one possible path; similar results can be achieved using other internal functions. The complete \ac{PoC}, demonstrating full arbitrary code execution, is available on GitHub~\cite{Digregorio-cve-2025-8747-anon}.
\texttt{Lambda} layers in Keras can be used not only to serialize Python bytecode but also to reference functions from specified Python modules. 
We found that an attacker can achieve arbitrary code execution through \emph{code reuse} by leveraging \texttt{Lambda} layers to abuse legitimate functions from Keras’s internal modules, effectively bypassing the restriction introduced after \hyperref[keras_first]{KV.\ref*{keras_first}} disclosure. As a \ac{PoC}, we crafted a model that disables \texttt{safe\_mode} during loading, even if it was initially enabled by the user, by invoking an internal Keras utility. Then, by relying on other internals, such as the function used to load the model (now with \texttt{safe\_mode} disabled), we show how to achieve full code execution. Our \ac{PoC} represents just one possible execution path; different exploits can be built using other internal functions (execution gadgets). A demonstrative snippet is provided in the appendices (Listing~\ref{fig:disable-safe-mode}), while the complete \ac{PoC} is available on GitHub.

%\sk{no enumerate qui}

% Its main steps are:
% \begin{enumerate}
%     \item Disable the global \texttt{safe\_mode} flag by calling Keras’s internal utility for setting global attributes.
%     \item Load a second model via \texttt{load\_model} (now unrestricted), pointing to a path under attacker control.
%     \item (Optional) Use \texttt{get\_file} to fetch a remote model, as also independently shown by JFrog researcher Andrey Polkovnichenko~\cite{PolkovnichenkoKerasSafeMode}. Alternatives include loading local files, bundling payloads in the original archive, or referencing external sources (e.g., \texttt{hf://}).
% \end{enumerate}

\mypar{Disclosure}
% We disclosed the vulnerability through a responsible disclosure process. Our report was considered a contemporary independent report of that by JFrog researcher Andrey Polkovnichenko~\cite{PolkovnichenkoKerasSafeMode}, which demonstrated the threat of abusing \texttt{Lambda} layers for arbitrary file downloading (while we showed \texttt{Lambda} abuse for arbitrary execution through gadget chains). The vulnerability was assigned the identifier \texttt{CVE-202X-XXX2}~\cite{CVE-2025-8747-anon}.
% The fix extends the validation checks introduced after the previous vulnerability by enforcing that any object accessed from an imported module must be an instance of \texttt{KerasSaveable}. Additionally, several internal Keras utilities that could be abused were blocklisted, including the ones we used in our \ac{PoC}.
We disclosed the vulnerability through a coordinated disclosure process. Two distinct CVEs were assigned to this vulnerability. The first (\texttt{CVE-2025-8747}, CVSS 8.6, CNA: Google LLC~\cite{CVE-2025-8747}) refers to the threat of arbitrary file download through specific gadget reuse, which we used as an optional step in our complete \ac{PoC}, with our report considered a contemporary independent report of that by JFrog researcher Andrey Polkovnichenko~\cite{PolkovnichenkoKerasSafeMode}. The second (\texttt{CVE-2025-9906}, CVSS 8.6, CNA: Google LLC~\cite{CVE-2025-9906}), instead, concerns code reuse to disable \texttt{safe\_mode}, thereby covering the broader and more severe threat of arbitrary code execution.
% Our report was considered a contemporary independent report of that by JFrog researcher Andrey Polkovnichenko~\cite{PolkovnichenkoKerasSafeMode}, which demonstrated the threat of abusing \texttt{Lambda} layers for arbitrary file downloading (while we showed \texttt{Lambda} abuse for arbitrary execution through gadget chains). The vulnerability was assigned the identifier \texttt{CVE-2025-8747}~\cite{CVE-2025-8747}.
The fix extends the validation checks introduced after \hyperref[keras_first]{KV.\ref*{keras_first}} by enforcing that any object accessed from an imported module must be an instance of \texttt{KerasSaveable}. Additionally, several internal Keras utilities that could be abused were blocklisted, including the ones we used in our \ac{PoC}.

\vksec{Silent Bypass via Legacy HDF5 Format}
\label{keras_third}
% \begin{marcosbox}
% \linkedtakeaways{TK.1 – Misplaced Trust in Format-Based Security; TK.1 – Misplaced Trust in Format-Based Security}
% \end{marcosbox}

% As anticipated in Section~\ref{sec:keras_sharing}, Keras continues to support loading legacy models in the HDF5 format. Due to their legacy nature, the security checks introduced for the \texttt{.keras} format do not apply to HDF5 models. To quote the official documentation: \emph{"This argument [\texttt{safe\_mode}] is only applicable to the Keras v3 model format"}~\cite{KerasSavingLoading}, which means there is no intended mechanism to restrict the content of \texttt{Lambda} layers in models stored in the HDF5 format.
% However, we observed that when an HDF5 model is loaded using \texttt{load\_model(..., safe\_mode=True)}, the \texttt{safe\_mode} flag is silently ignored—without any warning or error, even though such feedback would be reasonably expected given the impossibility of enforcing this mode. Technically, the argument is never forwarded to the internal legacy loading routine and therefore has no effect.
% In this case, no sophisticated techniques are required for an attack, as the legacy format permits deserialization of arbitrary code through unrestricted \texttt{Lambda} layers. An attacker can exploit the misleading behavior of \texttt{safe\_mode} with legacy formats, leveraging the fact that users may blindly trust
% the presence of a flag labeled as ``safe.''

As noted in Section~\ref{sec:keras_sharing}, Keras continues to support loading legacy models in the HDF5 format. Because of their legacy nature, some security checks introduced for the \texttt{.keras} format do not apply to HDF5 models. Specifically, there is no mechanism to restrict the content of \texttt{Lambda} layers~\cite{KerasSavingLoading}. However, we observed that when an HDF5 model is loaded using \texttt{load\_model(..., safe\_mode=True)}, the \texttt{safe\_mode} flag is silently ignored—without any warning or error, even though such feedback would be reasonably expected given the impossibility of enforcing this mode. Technically, the argument is never forwarded to the internal legacy loading routine and therefore has no effect. In this case, no sophisticated techniques are required for an attack, as the legacy format permits deserialization of arbitrary code through unrestricted \texttt{Lambda} layers. An attacker can then exploit this misleading behavior, leveraging the fact that users may blindly trust the presence of a flag labeled as ``safe.'' The \ac{PoC} is available on GitHub.

%not reasonably be expected to understand the underlying implementations or security guarantees of different model formats—trusting instead 
% {\color{blue}are (somehow) expected to understand the underlying lack of security guarantees derived from using different model formats, and cannot simply trust the presence of a flag labeled as ``safe.'' }

% An attacker can exploit this weakness by sharing a malicious HDF5 model file, which may be loaded by users even under the most secure settings. No sophisticated techniques are required, as the legacy format bypasses all security checks and inherently permits deserialization of insecure code.

% While the Google Security Team acknowledged this as an intentional design choice during disclosure, the behavior is misleading in practice—especially considering that users cannot reasonably be expected to distinguish between model formats. The Keras team agreed with this concern and committed to introducing an API change.

\mypar{Disclosure}
As in the previous cases, we disclosed the issue to the Google Open Source Security Team, which serves as a coordination channel for Google-affiliated open-source projects. 
However, this time our issue was marked as \textit{``Won’t Fix (Infeasible).''}
After a constructive discussion, Google Security clarified that they \textit{``won't treat Keras \texttt{safe\_mode}  as a security boundary''} anymore and that they \textit{``just don't think \texttt{safe\_mode}  is reliable enough to be a security boundary''}, further explaining that \textit{``the panel changed its view on this issue in [\hyperref[keras_second]{KV.\ref*{keras_second}} issue]''} and stating: \textit{``Don’t rely on \texttt{safe\_mode} (maybe a poor name) for that level of protection.''}
Moreover, they also clarified that \textit{``we don't speak for the Keras team—they might see it differently. If you want to see code changes in Keras for patching these issues, GitHub's the place to make it happen''}~\cite{GoogleSecurityTeam2025}.
Following this upstream referral, we then contacted the Keras team through GitHub's private advisory. 
Keras acknowledged the concern and fixed the issue in Keras version 3.11.3.
The fix extends \texttt{safe\_mode} to the legacy file format as well: the flag is now forwarded to legacy loading, ensuring that \texttt{Lambda} layers are constrained in the same way, including the additional restrictions introduced for \hyperref[keras_second]{KV.\ref*{keras_second}}. The vulnerability has been assigned the identifier \texttt{CVE-2025-9905} (CVSS 7.3, CNA: Google LLC)~\cite{CVE-2025-9905}.
% At the time of writing, the \texttt{safe\_mode} flag naming and related documentation have not been updated to eventually reflect Google’s revised position.}

% As in the other cases, we disclosed the issue to the Google Security Team. However, this time it was marked as \textit{“Won't Fix (Infeasible).”} Google Security clarified that they \textit{"won't treat Keras safe\_mode as a security boundary"} anymore, stating: \emph{“Don’t rely on \texttt{safe\_mode} (maybe a poor name) for that level of protection”}~\cite{GoogleSecurityTeam2025}.
% We then contacted the Keras team directly, who acknowledged the concern and fixed the issue in Keras version 3.11.3.%~\footnote{We are awaiting details on public disclosure}.
% %\sk{queste disclosuree ogni tanto suonano troppo come un racconto più che un discorso formale. Questa parentesi la sposterei in un footnote nel caso.}
% %\iono{la prentesi è per i reviewer, nel camera ready non ci sarà. Vedi footnote 1 come incipit}
% The fix extends \texttt{safe\_mode} to the legacy file format as well: the flag is now forwarded to legacy loading, ensuring that \texttt{Lambda} layers are constrained in the same way, including the additional restrictions introduced for \hyperref[keras_second]{KV.\ref*{keras_second}}. {\color{blue}The vulnerability has been assigned the identifier \texttt{CVE-2025-9905}~\cite{CVE-2025-9905}.}

\subsection{Skops}

\vssec{Abusing \texttt{MethodNode}}% for Attribute Traversal}
\label{skops_first}

In a Skops model, the \texttt{MethodNode} allows access to Python object attributes using dot notation. However, shortcomings in its design allow for traversal of the object graph of legitimate objects and access to sensitive Python internals. These can then represent powerful primitives, ultimately enabling arbitrary code execution during model loading. For example, a legitimate object may be instantiated using an \texttt{ObjectNode}, which enforces type validation and permits only trusted or explicitly allowed types. Once the object is in memory, however, an attacker can chain multiple \texttt{MethodNode} entries to traverse the object graph and access runtime structures such as \texttt{\_\_builtins\_\_}, which exposes dangerous functions like \texttt{eval} and \texttt{exec}. Furthermore, while the \texttt{\_\_class\_\_} and \texttt{\_\_module\_\_} fields of a \texttt{MethodNode} are validated when \texttt{get\_untrusted\_types()} and \texttt{load()} are called, the \texttt{func} field determines which attribute is accessed, allowing malicious attribute traversals to go unnoticed. 

During our analysis, we achieved arbitrary code execution using an apparently benign type returned by \texttt{get\_untrusted\_types()}, such as \texttt{builtins.int}. The specific type is irrelevant, as it is only checked during validation and never used by the Skops loading logic; in practice, any string can be substituted without affecting the outcome. A representative \texttt{schema.json} snippet from the \texttt{.skops} file of our \ac{PoC} is provided in the appendices (Listing~\ref{fig:methodnode-chain}), and the full \ac{PoC} is available on GitHub.

\mypar{Disclosure}
We disclosed the vulnerability to the Skops team via GitHub's private advisory and actively collaborated with the maintainers to develop and validate a mitigation. The fix was included in Skops version 0.12.0. The patch enforces that the \texttt{\_\_module\_\_} and \texttt{\_\_class\_\_} entries match those of the actual object passed to the \texttt{MethodNode}. In addition, it extends the untrusted types reported to the user by including any attribute accessed via \texttt{MethodNode}—that is, the concatenation \texttt{\_\_module\_\_.\_\_class\_\_.func}—thereby enabling human validation of the \texttt{func} entry as well. 
The vulnerability has been assigned the identifier \texttt{CVE-2025-54413} (CVSS 8.7, CNA: GitHub, Inc.)~\cite{CVE-2025-54413}.

\vssec{Bypassing Validation via \texttt{OperatorFuncNode}}
\label{skops_second}

The \texttt{OperatorFuncNode} allows invoking methods from Python’s \texttt{operator} module. However, similar to \hyperref[skops_first]{SV.\ref*{skops_first}}, we observed a mismatch between what is validated by \texttt{get\_untrusted\_types()} and \texttt{load()} and what is internally used by Skops during model loading, enabling unnoticed access to operator methods. Specifically, while the concatenation of the \texttt{\_\_module\_\_} and \texttt{\_\_class\_\_} fields is validated, the \texttt{\_\_module\_\_} value is ignored and only \texttt{\_\_class\_\_} is used. In practice, if \texttt{\_\_class\_\_} is set to the string “some\_method”, the function actually invoked is \texttt{operator.some\_method}, regardless of the value of \texttt{\_\_module\_\_} .This allows an attacker to supply a misleading, seemingly benign module path that passes validation, while the executed function is actually taken from the \texttt{operator} module. This ultimately enables code execution through methods such as \texttt{operator.call}, which invoke arbitrary targets with attacker-controlled arguments. A \texttt{schema.json} fragment showing the attack core is provided in the appendices (Listing~\ref{fig:operatorfuncnode}), while the \ac{PoC} demonstrating arbitrary code execution is available on GitHub.

\mypar{Disclosure}
We disclosed the vulnerability to the Skops team via GitHub's private advisory system and collaborated with the maintainers for remediation. The issue was resolved in Skops version 0.12.0 by enforcing the \texttt{\_\_module\_\_} field of an \texttt{OperatorFuncNode} to be set to "operator".
The identifier \texttt{CVE-2025-54412} (CVSS 8.7, CNA: GitHub, Inc.) has been assigned~\cite{CVE-2025-54412}.

% \begin{listing}[t]
% \centering
% \begin{minipage}{\linewidth}
% \begin{minted}[fontsize=\small, frame=single, bgcolor=gray!5, breaklines=true]{python}
% from skops.card import Card

% card = Card("model.skops")
% clf  = card.get_model()
% \end{minted}
% \caption{PoC showing that \texttt{Card.get\_model()} loads arbitrary files using \texttt{joblib} if not a valid ZIP in Skops.}
% \label{fig:card-joblib-bypass}
% \end{minipage}
% \end{listing}

\vssec{Silent Fallback to joblib in Model Card}
\label{skops_third}
% \begin{marcosbox}
% \linkedtakeaways{TK.1 – Misplaced Trust in Format-Based Security; TK.1 – Misplaced Trust in Format-Based Security}
% \end{marcosbox}
As previously noted in Section~\ref{sec:scikit}, Skops is developed with integration into the Hugging Face ecosystem in mind. To support this, Skops provides an API for generating model cards, which serve as structured documentation for models and include fields such as description, authors, diagrams, etc.~\cite{SkopsModelCard}.
% Once a \texttt{Card} object is created—typically specifying the model file and, optionally, a list of trusted types—the user can invoke \texttt{Card.get\_model()} to load the associated model.
When a \texttt{Card} object is created—typically specifying the model file and, optionally, a list of trusted types—Skops internally invokes \texttt{Card.get\_model()} to load the associated model.
If the provided model file is in the \texttt{.skops} format (i.e., a ZIP archive), the standard Skops \texttt{load()} function is used, applying all the security checks already discussed. 
% {\color{red}In particular, if the model includes any untrusted types and no corresponding trusted list is provided, an error is raised, consistent with Skops’ security model.
%The issue arises when the provided file is not a valid ZIP archive. In this case}
However, if the provided file is not a valid ZIP archive, the fallback mechanism silently switches to using \texttt{joblib.load()} to deserialize the model, without warning the user. Nevertheless, joblib does not provide the same protections as Skops and allows arbitrary code execution via pickle-based deserialization.
Importantly, this behavior is based on the file's actual format—not its extension—making it especially difficult for users to detect. An example \ac{PoC} is available on GitHub.

\mypar{Disclosure}
We disclosed the vulnerability to the Skops team via GitHub's advisory system and proposed a fix to the maintainers, which was accepted and included in Skops version 0.13.0. The fix disallows the use of joblib unless explicitly authorized by the user through the new \texttt{allow\_pickle} argument during \texttt{Card} creation.
The vulnerability has been assigned the identifier \texttt{CVE-2025-54886} (CVSS 8.4, CNA: GitHub, Inc.)~\cite{CVE-2025-54886}.

\subsection{Hugging Face}
\begin{table}[t]
\centering
\caption{Detection results of Hugging Face scanning tools for our \acp{PoC} and baselines, as presented in the interface.
}
\label{tab:scan-results}
\resizebox{\linewidth}{!}{%
\begin{tabular}{c|cccc|c}
\toprule
\textbf{Test (format)} & \textbf{Picklescan} & \textbf{ClamAV} & \textbf{Protect AI} & \textbf{JFrog} & \textbf{Final label} \\
\midrule
\hyperref[keras_first]{KV.\ref*{keras_first}} (.keras)        & not a pickle & No issue & Unsafe & No Issue & Unsafe \\
\hyperref[keras_second]{KV.\ref*{keras_second}} (.keras)      & not a pickle & No issue & Suspicious & (**)  & (**) \\
\multirow{2}{*}{\shortstack{\hyperref[keras_third]{KV.\ref*{keras_third}} (HDF5) \\ (same as M-L)}} 
                   & \multirow{2}{*}{not a pickle} 
                   & \multirow{2}{*}{No issue}
                   & \multirow{2}{*}{ (*) }
                   & \multirow{2}{*}{Unsafe}
                   & \multirow{2}{*}{Unsafe} \\
& & & & & \\
\hyperref[skops_first]{SV.\ref*{skops_first}} (.skops)      & not a pickle & No issue & No issue & not a model & Safe \\
\hyperref[skops_second]{SV.\ref*{skops_second}} (.skops)      & not a pickle & No issue & No issue & not a model & Safe \\
\hyperref[skops_third]{SV.\ref*{skops_third}} (pickle)      & not a pickle & No issue & Unsafe & Unsafe & Unsafe \\
\midrule
B-L (HDF5)         & not a pickle & No issue & Suspicious & Unsafe & Unsafe \\
B-L (.keras)       & not a pickle & No issue & Suspicious & Unsafe & Unsafe \\
M-L (.keras)       & not a pickle & No issue & No issue & Unsafe & Unsafe \\
NL (HDF5)          & not a pickle & No issue & No issue & No issue & Safe \\
NL (.keras)        & not a pickle & No issue & No issue & No issue & Safe \\
\bottomrule
\end{tabular}%
\vspace{0.4em}
}
{\raggedright\scriptsize
(*) The scanning tool did not return any results (i.e., an empty label). \par
(**) The scanning tool remained stuck in the “Queued” status. No final label was therefore computed. Re-uploading the model produced the same result. \par
Legend: B = Benign, M = Malicious, L = \texttt{Lambda}, NL = No \texttt{Lambda}. \par
}
\end{table}

% \begin{table*}[t]
% \centering
% \caption{Detection results of Hugging Face scanning tools, as displayed in the interface.
% B = Benign, M = Malicious, L = Lambda, NL = No Lambda.}
% \label{tab:scan-results-rotated}
% \resizebox{\linewidth}{!}{%
% \begin{tabular}{c|ccccccccccc}
% \toprule
% \textbf{Tool} & \textbf{KV.1} & \textbf{KV.2} & \textbf{KV.3} & \textbf{SV.1} & \textbf{SV.2} & \textbf{SV.3} & \textbf{B-L (HDF5)} & \textbf{B-L (.keras)} & \textbf{M-L (.keras)} & \textbf{NL (HDF5)} & \textbf{NL (.keras)} \\
% \midrule
% Picklescan & not a pickle & not a pickle & not a pickle & not a pickle & not a pickle & not a pickle & not a pickle & not a pickle & not a pickle & not a pickle & not a pickle \\
% ClamAV     & No issue & No issue & No issue & No issue & No issue & No issue & No issue & No issue & No issue & No issue & No issue \\
% Protect AI & Unsafe & Suspicious & - & No issue & No issue & Unsafe & Suspicious & Suspicious & No issue & No issue & No issue \\
% JFrog      & No Issue &  & Unsafe & not a model & not a model & Unsafe & Unsafe & Unsafe & Unsafe & No issue & No issue \\
% HF label   & Unsafe &  & Unsafe & Safe & Safe & Unsafe & Unsafe & Unsafe & Unsafe & Safe & Safe \\
% \bottomrule
% \end{tabular}%
% }
% \end{table*}

Following the threat model described in Section~3.1, we assess whether an attacker can distribute exploits that target framework-level vulnerabilities triggered during model loading, without being detected by the scanning tools integrated into model hubs.
In other words, we assess whether these scanners provide an effective additional line of defense when framework-level protections fail.
To do so, we tested the exploits for vulnerabilities we identified in Keras and Skops against the scanners integrated into Hugging Face, which is the only major hub that integrates scanners. We uploaded the original \acp{PoC} for each vulnerability without obfuscation or modification. Each exploit opens \texttt{/bin/sh}.
% We evaluate Hugging Face security by testing the ability of its integrated scanning tools to detect the exploit for vulnerabilities we identified in Keras and Skops. We uploaded the original \acp{PoC} for each vulnerability without obfuscation or modification. Each exploit opens \texttt{/bin/sh}.
As a baseline, we also uploaded a set of additional Keras models: one containing a benign \texttt{Lambda} layer (doing nothing), one containing a \textit{malicious} \texttt{Lambda} layer calling \texttt{/bin/sh}, and one with no \texttt{Lambda} layers. For each of these, we uploaded both the HDF5 and \texttt{.keras} versions\footnote{It is worth noting that the baseline model with a \texttt{Lambda} layer calling \texttt{/bin/sh} in HDF5 format coincides exactly with the \ac{PoC} of \hyperref[keras_third]{KV.\ref*{keras_third}}, since the exploit consists of using a legacy HDF5 model.}. Our experiments were performed several weeks after the CVEs were publicly disclosed, assessing the capability of scanners to detect public threats replicable by attackers. 

We made all loaded artifacts (or the corresponding generation scripts) publicly available. Before uploading any malicious models, we requested and obtained explicit permission from Hugging Face. The results are shown in Table~\ref{tab:scan-results}.

\mseff{Effectiveness of Hub-Integrated Model Scanners}\label{ms}

%The scan results are reported in Table~\ref{tab:scan-results}. 

\mypar{Picklescan}  
As evident from Table~\ref{tab:scan-results}, this scanner does not detect any of our \acp{PoC}, classifying every file as "not a pickle," including the malicious pickle of \hyperref[keras_third]{KV.\ref*{keras_third}}. 
The reason lies in the extension chosen for the PoC (\texttt{.skops}). To confirm it, we uploaded the same pickle file under different names. Files with extensions such as \texttt{.h5}, \texttt{.onnx}, \texttt{.pkl}, and \texttt{.pt} were flagged as (malicious) pickle, while \texttt{.json}, \texttt{.skops}, or \texttt{.keras} were flagged as "not a pickle".

% \mypar{HF PickleScan}  
% As evident from the results, this scanner is not helpful for any of our \acp{PoC}, classifying every file as ``not pickle.'' While this outcome is expected for most of them, it is surprising in the case of the KV.3 \ac{PoC}, which is in fact a malicious pickle. The reason behind this misclassification lies in the filename chosen for the PoC, which ended with the extension \texttt{.skops}. To confirm this, we uploaded the same pickle file (executing \texttt{/bin/sh}) under different names. Interestingly, it was recognized as a pickle, scanned, and flagged as ``unsafe'' when given extensions such as \texttt{.h5}, \texttt{.onnx}, \texttt{.pkl}, and \texttt{.pt}, but it was not recognized as a pickle when named with extensions such as \texttt{.json}, \texttt{.skops}, or \texttt{.keras}. {\color{red} This highlights a serious concern.}{\color{blue} LOL.}
% \sk{jk, ma toglierei l'ultima frase, non aggiunge niente se non un "coglioni" neanche troppo velato.}

\mypar{ClamAV}  
All model scans reported “No issue.” This result is expected, as ClamAV is a general-purpose malware scanner and is not designed to detect \ac{ML} framework-level threats.
%, such as those in our \acp{PoC}.

\mypar{Protect AI}  
% Protect AI defines a set of threats related to model sharing. Its scanner, called Guardian, checks files against these definitions and produces a report accessible through Hugging Face of the detected threats. 
% \iono{this was expected since there is an annoincment of hf on the cve...}
% Protect AI returns matches for KV.1, KV.2, and two baselines: for KV.1, the \ac{PoC} matched the threat “PAIT-KERAS-301: Keras Model Custom Layer Detected at Model Run Time,” which explicitly covers cases where non-Keras layers are included within a Keras model. This corresponds exactly to the basic \ac{PoC} we uploaded for KV.1, confirming the tool’s ability to identify such cases. For KV.2, the suspicious match was with the threat “PAIT-KERAS-100: Keras Model Lambda Layer Can Execute Code At Load Time.” As demonstrated by our baselines, however, this is a generic threat that detects the presence of Lambda layers, without evaluating their operations. In fact, benign Lambda layers trigger the same warning, generating a false positive that may lower the user's alert level when receiving a positive scan. Interestingly, the baseline containing the Lambda layer calling \texttt{/bin/sh} in \texttt{.keras} was not flagged, while the equivalent in \texttt{.h5} (KV.3) produced no label at all. We regard these as a false negative and a malfunction, respectively. 
% Overall, this test highlight concerns about the reliability of this threat definition.
Protect AI’s Guardian~\cite{ProtectAIGuardian} scanner defines a set of model-sharing threats~\cite{ProtectAIThreats}. When a model is uploaded to Hugging Face, the scanner checks the files against these definitions and generates a report that alerts users if any threats are detected. In our test, this scanner returned matches for \hyperref[keras_first]{KV.\ref*{keras_first}}, \hyperref[keras_second]{KV.\ref*{keras_second}}, and two baseline models. 

For \hyperref[keras_first]{KV.\ref*{keras_first}}, the \ac{PoC} triggered the threat “\textit{PAIT-KERAS-301: Keras Model Custom Layer Detected at Model Run Time,”} which explicitly covers cases where non-Keras layers are included within a Keras model. This corresponds exactly to the basic \ac{PoC} we uploaded, confirming the tool’s ability to identify such cases. This result was somewhat expected, as following the publication of the CVE associated with \hyperref[keras_first]{KV.\ref*{keras_first}}, Hugging Face announced new Protect AI threat definitions, explicitly citing our CVE as an example of the type of threat detectable through those updates~\cite{MorganPAI6Months}.
\hyperref[keras_second]{KV.\ref*{keras_second}} was instead flagged as “suspicious” under the threat \textit{“PAIT-KERAS-100: Keras Model Lambda Layer Can Execute Code At Load Time.”} However, as demonstrated by our baseline models, this threat definition is generic and triggered by the presence of \texttt{Lambda} layers, rather than their internal operations. 
As a result, benign \texttt{Lambda} layers also trigger this warning, leading to false positives. This may, in turn, reduce users' perceived severity due to alert fatigue.
Interestingly, the baseline containing a \texttt{Lambda} layer that invokes \texttt{/bin/sh} in a \texttt{.keras} file was not flagged (i.e., a false negative), while the equivalent model in \texttt{.h5} format (\hyperref[keras_third]{KV.\ref*{keras_third}}) did not produce any label. Overall, this test raises concerns regarding the accuracy of this particular threat definition.

Regarding the Skops models, the pickle file for \hyperref[skops_third]{SV.\ref*{skops_third}} was correctly detected, as expected for a standard (malicious) pickle file. However, neither of the \texttt{.skops} files for \hyperref[skops_first]{SV.\ref*{skops_first}} or \hyperref[skops_second]{SV.\ref*{skops_second}} raised any issues. This is consistent with the fact that no threats are explicitly defined for Skops in Guardian. However, it is concerning that these cases were flagged as “No issue” rather than returning a more neutral (and informative) label indicating a lack of compatibility.

\mypar{JFrog}
% JFrog correctly classified the baseline Keras models containing a malicious \texttt{Lambda} as “Unsafe,” and the non-malicious Keras models without \texttt{Lambda} layers as “No issue.” It also successfully identified the malicious pickle file of \hyperref[skops_third]{SV.\ref*{skops_third}}.
% However, similar to Protect AI, it misclassified Keras models with benign \texttt{Lambda} layers as “Unsafe”—an even stronger flag than the one assigned by Protect AI—thus producing false positives. The reported details cited \textit{“models with Lambda layers containing malicious code,”} which did not align with reality.
% JFrog also failed to detect our exploit for \hyperref[keras_first]{KV.\ref*{keras_first}}, demonstrating difficulties in identifying newer Keras threats, and provided no decision for \hyperref[keras_second]{KV.\ref*{keras_second}}, which remained stuck in the status “Queued” despite being uploaded to the same repository and at the same time as the others. Re-uploading \hyperref[keras_second]{KV.\ref*{keras_second}} yielded the same result. Finally, both \texttt{.skops} models (\hyperref[skops_first]{SV.\ref*{skops_first}} and \hyperref[skops_second]{SV.\ref*{skops_second}}) were labeled as “not a model,” a classification that can be explained by the lack of support for this format, but which may be misleading for the end user.
JFrog correctly classified the baseline Keras models containing a malicious \texttt{Lambda} as “Unsafe,” and the non-malicious Keras models without \texttt{Lambda} layers as “No issue.” It also successfully identified the malicious pickle file of \hyperref[skops_third]{SV.\ref*{skops_third}}.
However, similar to Protect AI, it misclassified Keras models with benign \texttt{Lambda} layers as “Unsafe”—an even stronger flag than the one assigned by Protect AI—thus producing false positives. The reported details cited \textit{“models with Lambda layers containing malicious code,”} which did not align with reality.
JFrog also failed to detect our exploit for \hyperref[keras_first]{KV.\ref*{keras_first}}, demonstrating difficulties in identifying newer Keras threats, and provided no decision for \hyperref[keras_second]{KV.\ref*{keras_second}}, which remained stuck in the status “Queued” despite being uploaded to the same repository and at the same time as the others. Re-uploading \hyperref[keras_second]{KV.\ref*{keras_second}} yielded the same result. Finally, both \texttt{.skops} models (\hyperref[skops_first]{SV.\ref*{skops_first}} and \hyperref[skops_second]{SV.\ref*{skops_second}}) were labeled as “not a model”. While this outcome can be explained by the lack of support for the \texttt{.skops} format, it may nevertheless be misleading for end users.

\mypar{Final Label}
The final label shown by Hugging Face corresponds to the most severe label assigned by its scanners. This approach allows threats such as the PoC for \hyperref[keras_first]{KV.\ref*{keras_first}} (detected only by Protect AI) to be flagged as “Unsafe,” thereby compensating for false negatives. On the other hand, this strategy also amplifies false positives: harmless Keras models with \texttt{Lambda} layers are escalated to “Unsafe.”
Concerningly, since none of the scanners support Skops models, the final label for both \hyperref[skops_first]{SV.\ref*{skops_first}} and \hyperref[skops_second]{SV.\ref*{skops_second}} PoCs was “Safe.” This represents a clear false negative and is particularly problematic, as a reassuring label such as “Safe” was assigned despite the absence of compatible scanners.
%\section{User Perception: A Survey on Model Sharing}
\section{Survey on User Perception (RQ3)}
\label{sec:user-perception}

To assess whether the narratives promoted by certain frameworks and hubs influence \ac{ML} practitioners’ perceptions, we conducted a public survey.
The survey was distributed via social media and direct outreach to professionals in both academic and industrial \ac{ML} communities. To minimize bias, we did not disclose the cybersecurity focus of the study. This enables gathering more authentic insights into the natural concerns and mental models that participants associate with model sharing. Responses were anonymous, and no sensitive data was collected. 
The survey included 14 multiple-choice questions, 6 of which allowed optional open-ended answers. It was structured into three main parts, presented in the following sections.
All questions, raw results, and scripts used for both statistics extraction and plotting are publicly available.
% % in text
% \begin{enumerate}[label={}, labelsep=0pt, labelwidth=0pt,
%                   leftmargin=0pt, itemindent=0pt,
%                   itemsep=0pt, topsep=0pt]
%   \item \textbf{UP.1 - Demographics:} Participants reported machine learning expertise (1–5), area of work, and prior experience loading shared \ac{ML} models.
%   \item \textbf{UP.2 - Perception of Model Loading Security:} Participants rated comfort (1–10) running short Keras/PyTorch loading snippets under varied configs (e.g., with/without \texttt{safe\_mode}) and selected any concerns.
%   \item \textbf{UP.3 - Hubs Influence:} Participants indicated whether security tools/scanning on model hubs (e.g., Hugging Face Hub) would affect their concerns about loading shared models.
% \end{enumerate}

\mypar{Limitations}
% While the survey provides valuable insight, the number of participants was relatively limited. As a result, the findings should be interpreted as indicative rather than representative of the entire \ac{ML} community. Nonetheless, the results reveal meaningful trends in the model sharing security perception.
%in the influence of the API design on the perceived security of model loading workflows, highlighting how certain configurations can strongly shape 
While the survey offers insights, its limited number of participants means the findings should be considered as indicative rather than representative of the broader \ac{ML} community. Nevertheless, the results highlight meaningful trends in how model sharing security is perceived. To assess the robustness of these observations, we complement the analysis with statistical tests evaluating the significance of the reported outcome.

\begin{figure}[t]
\centering
\includegraphics[width=0.9\linewidth]{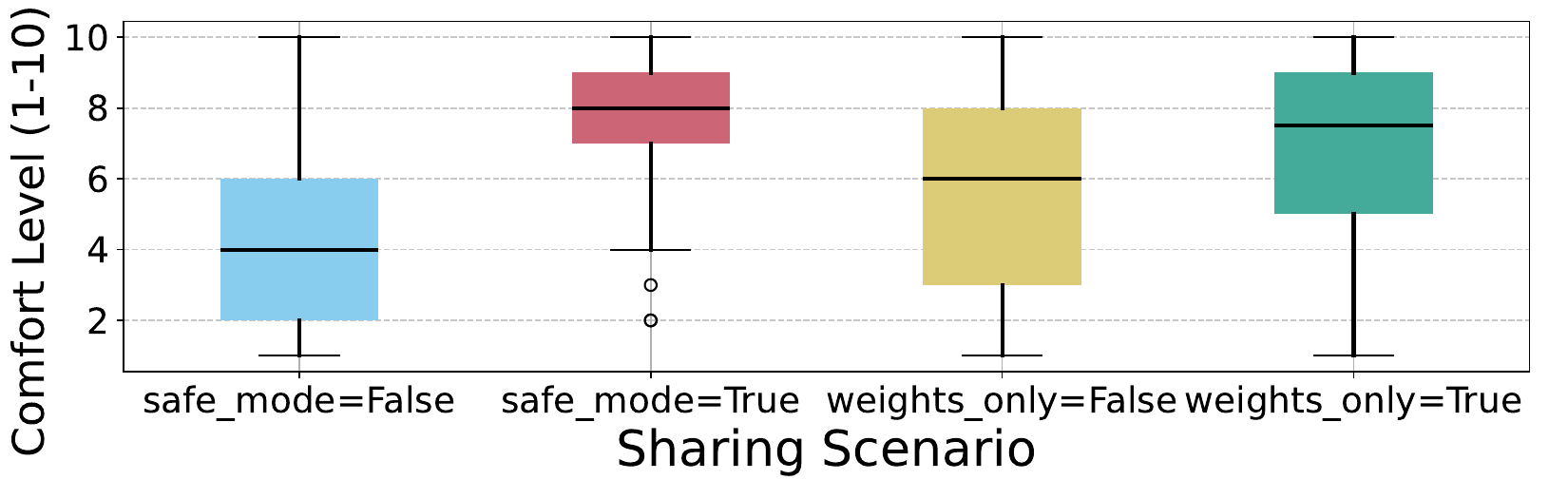}
\caption{Distribution of user comfort levels (1–10) when loading shared models under different configurations.}
\label{fig:safe_mode_perception}
\end{figure}

\upsec{Demographics}
\label{UP1}

%A total of 62 participants completed the survey, 53 (85.5\%) of whom reported experience loading or sharing \ac{ML} models; our analysis focuses on this group.
A total of 62 participants completed the survey. Among them, 53 (85.5\%) reported experience loading or sharing \ac{ML} models. All results presented in the following analysis are restricted to these 53 participants.

Within this group, 33 (62.3\%) listed \ac{ML} or \acf{AI} as their primary area of expertise, 9 (17.0\%) selected cybersecurity, and the remaining 11 (20.7\%) came from related fields such as data science, software engineering, high-performance computing, or robotics. 
The average self-assessed expertise in \ac{ML} was 3.47 on a scale from 1 (basic familiarity) to 5 (expert).

\upsec{Perception of Model Loading Security}
\label{UP2}
To assess participants’ perceived security when loading shared models, we asked them to rate their comfort with four Python scripts—two using Keras and two using PyTorch. These frameworks were selected because both expose explicitly security-oriented loading-time flags: \texttt{safe\_mode} in Keras and \texttt{weights\_only} in PyTorch. Together, they also cover a range of persistence strategies—self-contained vs.\ non-self-contained, and data-based vs.\ code-based formats—providing valuable contrast for our analysis.
For each case, participants were also asked to specify the reasons for any lack of comfort, selecting from predefined categories—such as ethical concerns, data bias, and cybersecurity risks—or by providing an open-ended response.

\mypar{Keras \texttt{safe\_mode}}
The survey included two Keras code snippets differing only in the value of the \texttt{safe\_mode} flag (\texttt{False} vs.\ \texttt{True}).
As shown in Figure~\ref{fig:safe_mode_perception}, average comfort levels grow from 4.5/10 with \texttt{safe\_mode=False} to 7.8/10 with \texttt{safe\_mode=True}, with responses shifting from widely dispersed to tightly clustered around higher scores—indicating stronger perceived security. 
% Concerns about arbitrary code execution dropped from 44 (83\%) participants with \texttt{safe\_mode=False} to just 8 (15.1\%) with \texttt{safe\_mode=True}, half of whom declared cybersecurity as their area of expertise.
The number of participants expressing concerns about arbitrary code execution decreased from 44 (83\%) to 8 (15.1\%), half of whom identified cybersecurity as their primary area of expertise.

% The first two code snippets in the survey showed identical Python code using Keras’s \texttt{load\_model()} method, differing only in the value of the \texttt{safe\_mode} parameter: once set to \texttt{True}, and once to \texttt{False}.

% Figure~\ref{fig:safe_mode_perception} shows the distribution of user comfort levels (rated from 1 to 10) under both settings. Notably, the average comfort level increased from \textbf{4.5/10} when \texttt{safe\_mode=False} to \textbf{7.8/10} when \texttt{safe\_mode=True}. Beyond the change in average, the figure also illustrates a marked shift in distribution. With \texttt{safe\_mode=False}, comfort levels are widely spread across the scale, indicating diverse and uncertain levels of trust. In contrast, when \texttt{safe\_mode=True}, responses cluster more tightly around higher values, reflecting a strong consensus in perceived security improvement. 

% Participants' stated concerns also varied significantly. The number of participants reporting concerns about arbitrary code execution dropped from \textbf{44} out of 53 (83\%) when \texttt{safe\_mode=False} to only \textbf{8} participants (15.1\%) when \texttt{safe\_mode=True}. Interestingly, 4 out of these 8 (50\%) had a background in cybersecurity.

\mypar{PyTorch \texttt{weights\_only}}
The survey included two PyTorch snippets, each setting the \texttt{weights\_only} flag to either \texttt{False} or \texttt{True}. In the latter case, participants were informed that model definitions must be provided separately, possibly by importing shared code.
Here, the shift in perceived security was smaller than in the previous case: the average comfort level grew from 5.9 to 7.0. As shown in Figure~\ref{fig:safe_mode_perception}, the dispersed distribution of responses indicates uncertainty among participants.
% However, concerns about arbitrary code execution dropped more sharply—though still less than in the Keras case—from 28 (52.8\%) to 13 (24.5\%), even though the \textit{more secure} option requires manually loading code for the model definition. 
However, the number of participants expressing concerns about arbitrary code execution decreased more sharply, though still less than in the Keras case, from 28 (52.8\%) to 13 (24.5\%), even though the \textit{more secure} option requires manually loading code for the model definition.
Notably, despite unrestricted pickle use, respondents reported moderate concern in the \texttt{weights\_only=False} setting.

\mypar{Statistical Validation of the Observed Effects}
To evaluate whether the observed differences in perceived security are statistically robust, we conduct a paired Wilcoxon signed-rank test~\cite{wilcoxon1945individual}. In this context, $p$ denotes the p-value of the test and $r$ the corresponding effect size. The results indicate a statistically significant increase in reported comfort when \texttt{safe\_mode=True} with a large effect size ($p = 3.14 \times 10^{-9}$, $\lvert r \rvert = 0.87$). A statistically significant comfort increase, though more moderate, was also observed when \texttt{weights\_only=True} ($p = 0.0054$, $\lvert r \rvert = 0.42$). These results indicate that the differences in perceived security reported in Figure~\ref{fig:safe_mode_perception} are unlikely to arise from random variation and suggest that the presence of security-oriented mechanisms meaningfully influences user perception of model loading security.

% The other two questions referenced the \texttt{weights\_only} mode of PyTorch, providing two code snippets with the option set to \texttt{False} and \texttt{True}, respectively. When \texttt{weights\_only=True}, we explicitly noted that the user must define and instantiate the model manually, potentially by loading external shared Python code.

% In this case, the difference in perceived security was less pronounced. The average comfort level increased from \textbf{5.9} (when \texttt{weights\_only=False}) to \textbf{7.0} (when \texttt{weights\_only=True}). Notably, the comfort level was relatively high from the beginning, despite the underlying use of raw pickle in the \texttt{False} setting. As shown in Figure~\ref{fig:safe_mode_perception}, the distribution of responses remained broad and relatively dispersed in both settings, indicating persistent uncertainty among participants. 

% On the other hand, participants' stated concerns showed stronger variation. The number of participants expressing concern about arbitrary code execution dropped from \textbf{28} (52.8\%) when \texttt{weights\_only=False} to \textbf{13} (24.5\%) when \texttt{weights\_only=True}, despite the need to load or import external Python code.

\upsec{Impact of Model-Sharing Hubs}
\label{UP3}
We asked participants whether their comfort level would change when loading models from hubs such as Hugging Face, which integrates various security scanners. In total, 39/53 participants (73.6\%) responded that the use of such hubs would \emph{increase} their level of comfort when loading shared models.

\subsection{Results Interpretation}
These results highlight a clear shift in user perception driven by the presence of specific flags in the model-loading function, particularly among participants who did not identify as cybersecurity experts.
% {\color{blue} Notably, the results indicate a stronger influence of \texttt{safe\_mode} compared to \texttt{weights\_only}. This discrepancy may stem from users perceiving the need to manually define and instantiate the model (required by \texttt{weights\_only=True}) as an additional source of risk. At the same time, factors such as flag naming contribute to shaping users’ sense of security. Indeed, the confidence with \texttt{weights\_only=False} is higher than with \texttt{safe\_mode=False}. In contrast, we observe the reverse pattern when both flags are \texttt{True}.}
Notably, the results indicate a stronger influence of \texttt{safe\_mode} compared to \texttt{weights\_only}. This discrepancy may stem from users perceiving the need to manually define and instantiate the model (required by \texttt{weights\_only=True}) as an additional source of risk. At the same time, it may also reflect how different flag-naming choices shape users’ perception of security.
Finally, the results confirm that hub-level scanning features also have a significant effect on user perception.

\section{Takeaways}

% So far, we have analyzed the actual security of model sharing across popular machine learning frameworks and platforms, particularly those that advertise themselves as security-oriented. In this section, we step back from individual vulnerabilities to examine the broader implications of our findings. Our goal is to reflect on how secure model sharing is
% \sk{Python based qua sotto va cambiato direi?}
% approached in Python-based ecosystems and what assumptions persist among developers, users, and maintainers.
%So far, we have analyzed the security posture of model sharing across popular \acl{ML} frameworks and hubs, and evaluated the actual guarantees offered by those that advertise themselves as security-oriented. 
%In this section, we take a broader perspective to discuss the wider implications of our findings. Our discussion is further enriched with guidelines for the community—targeted at both \ac{ML} practitioners and framework or hub maintainers—with the goal of raising awareness and promoting authentically informed decision-making regarding model sharing practices.
In this section, we take a broader perspective to discuss the implications of our findings. Our discussion is enriched with suggestions for the community, directed at both \ac{ML} practitioners and framework or hub maintainers, to raise awareness and enable sound security choices.

% \mypar{General Suggestions for the Community}
% Before proceeding, we briefly list a set of evergreen software security practices—rooted in decades of experience with code sharing and software repositories—that we believe would also benefit the \ac{ML} community during model sharing:
% \begin{itemize}[label=\textbullet, 
%                 labelsep=0.5em, % space between bullet and text
%                 leftmargin=*, 
%                 itemsep=0pt, 
%                 topsep=0pt]
%     \item Rely only on trusted publishers; if not feasible, carefully consider user feedback and community reputation.

%     \item Always cross-check the source of a shared model to prevent impersonation or spoofing.

%     %\item Take countermeasures against supply chain attacks, such as verifying hashes of models that are known to be trustworthy.
%     \item Take countermeasures against supply chain attacks, such as verifying model hashes against trusted references.

%     \item Execute shared models in isolated environments to mitigate potential damage from malicious behavior.
% \end{itemize}

\subsection{The Illusion of \textit{Secure} File Formats}
\begin{marcosboxgreyopen}
\textbf{Derived from:}
\hyperref[saf]{SA.F}, \hyperref[keras_first]{KV.\ref*{keras_first}}, \hyperref[keras_second]{KV.\ref*{keras_second}}, \hyperref[skops_first]{SV.\ref*{skops_first}}, \hyperref[skops_second]{SV.\ref*{skops_second}}
\end{marcosboxgreyopen}
%It is tempting to reduce the security of model sharing to the nature of the adopted file formats. 
Data-based formats are often perceived as \textit{more secure} than code-based formats, primarily because they do not directly serialize executable code. This belief is reinforced by the documentation and narrative of explicitly security-oriented frameworks.%, which frame the absence of pickle as a guarantee of security.
% While this claim may seem reasonable, our analysis in Section~\ref{sec:whensecisnotsec} debunks it by highlighting that the file format alone does not determine the security of model sharing, and that the reality is far more complex.
While this claim may seem reasonable, our analysis in Section~\ref{sec:whensecisnotsec} demonstrates that the file format alone does not determine the security of model sharing, and that the reality is far more complex.
In particular, both Keras and Skops use JSON to represent the logical structure of executable code. Security, therefore, depends on strict validation and restrictions applied during the translation from JSON to code objects, resulting in a threat model similar to that of code-based formats. This fragility was evident in our discovered vulnerabilities (\hyperref[keras_first]{KV.\ref*{keras_first}}, \hyperref[keras_second]{KV.\ref*{keras_second}}, \hyperref[skops_first]{SV.\ref*{skops_first}}, and \hyperref[skops_second]{SV.\ref*{skops_second}}) and the subsequent patches. Indeed, these fixes restricted the excessive flexibility and addressed missing checks in how code was reconstructed from JSON. In other words, model expressiveness impacts security regardless of the serialization format used: higher expressiveness increases the attack surface. Alternative approaches exist, such as constrained formats like ONNX~\cite{onnx}, which implicitly reduce the attack surface by limiting the set of allowed operators, albeit at the cost of flexibility (see Section~\ref{subsec:block_allow}).

\mypar{JSON Models Are Code}
To conclude this point, we report a statement from the Google Security Team during a private exchange related to one of our Keras disclosures:
% \begin{list}{}{\setlength{\leftmargin}{10pt}%
%                \setlength{\rightmargin}{10pt}%
%                \setlength{\topsep}{10pt}%
%                \setlength{\parsep}{10pt}%
%                \setlength{\partopsep}{0pt}}
% \item\textit{"Basically, a truly \texttt{safe\_mode} in Python for this isn't really possible. Loading untrusted models is like running untrusted code—models are code."}~\cite{GoogleSecurityTeam2025}
% \end{list}

\smallskip
\noindent
\textit{"Basically, a truly \texttt{safe\_mode} in Python for this isn't really possible. Loading untrusted models is like running untrusted code—models are code."}~\cite{GoogleSecurityTeam2025}
\smallskip

\begin{marcosboxgreyclosed}
\mypar{Suggestions for the Community}
Do not rely on file formats as a guarantee of security. What truly matters is what the format contains and how that content is processed. Sharing code objects is always inherently risky.
\end{marcosboxgreyclosed}

\subsection{Block or Allow: The Security Trade-Off}\label{subsec:block_allow}
\begin{marcosboxgreyopen}
\textbf{Derived from:}
\hyperref[saf]{SA.F}, \hyperref[keras_first]{KV.\ref*{keras_first}}, \hyperref[keras_second]{KV.\ref*{keras_second}}, \hyperref[skops_first]{SV.\ref*{skops_first}}, \hyperref[skops_second]{SV.\ref*{skops_second}}
\end{marcosboxgreyopen}

As with software in general, designing a secure sharing mechanism requires trade-offs between flexibility, usability, and security. Across frameworks, security is often enforced via either allowlists or blocklists. While these approaches reduce the attack surface, they inevitably limit flexibility, since anything outside an allowlist or inside a blocklist is either prohibited or left to human-based fallback mechanisms. %In the \ac{ML} domain, such restrictions can directly translate into reduced model performance on specific tasks.

ONNX~\cite{onnx}, although not explicitly designed for security, restricts models to a limited set of operators, enabling interoperability across frameworks and implicitly reducing the attack surface, but at the cost of flexibility, as only certain models and training capabilities are supported.

Methods claimed to be security-oriented follow a similar pattern when patching new vulnerabilities. In Keras, the fix to \hyperref[keras_first]{KV.\ref*{keras_first}} involved introducing an allowlist of Keras' modules to reconstruct the saved objects. However, as demonstrated in \hyperref[keras_second]{KV.\ref*{keras_second}}, this hardening could be bypassed through code reuse, requiring further restrictions and blocklisting.

% Weights-only formats also reflect this allowlisting-like pattern. PyTorch explicitly advertises its weights-only mode as security-oriented, since it loads only tensors from pickle and thereby enforces a narrow allowlist to prevent execution of arbitrary code. Keras, in contrast, also supports saving only weights but does not frame this as a security feature. In both cases, the attack surface is reduced, yet functionality is constrained, and the risk reappears when the model architecture code, required to restore the model object, comes from unverified sources.
PyTorch promotes its \texttt{weights\_only} mode as security-oriented, as it allows only tensors, enforcing a narrow type allowlist to prevent arbitrary code execution. Keras, in contrast, while also supporting the saving of weights only, does not present this as a security measure. In both cases, the attack surface is reduced, but functionality is constrained, shifting the problem to validating the trustworthiness of the sources providing the model architecture code.

Ultimately, in some cases, allowlisting and blocklisting propagate to users themselves, who act as the last line of defense. For instance, in Skops users must manually review and approve untrusted types. However, as we demonstrate in \hyperref[skops_first]{SV.\ref*{skops_first}} and \hyperref[skops_second]{SV.\ref*{skops_second}}, this mechanism is prone not only to human error but also to framework-level flaws, which may allow attackers to exploit user-allowlisted types while gaining capabilities for other, riskier types.
%These issues compound with inherent framework vulnerabilities that may still allow arbitrary code execution.

\begin{marcosboxgreyclosed}
\mypar{Suggestions for the Community}
Learn from decades of experience in trade-offs between flexibility and security. Allowlists offer strong security, but require continuous maintenance and can affect usability. Be aware and don’t rely on them alone: while they reduce the attack surface, they are not infallible, and should never be blindly trusted. Evaluate how strong limitations might offload security responsibilities to users, reintroducing risks.
\end{marcosboxgreyclosed}

\subsection{The Security Cost of Slow Adoption} 
\begin{marcosboxgreyopen}
\textbf{Derived from:}
\hyperref[keras_third]{KV.\ref*{keras_third}}, \hyperref[skops_third]{SV.\ref*{skops_third}}
\end{marcosboxgreyopen}
\hyperref[keras_third]{KV.\ref*{keras_third}} and \hyperref[skops_third]{SV.\ref*{skops_third}} give us hints about another systemic issue: the tradeoff between legacy compatibility and security. In both cases, legacy formats using pickle deserialization (e.g., HDF5 in Keras or joblib in scikit-learn) bypass the modern security mechanisms and allow trivial code execution, even when loaded under supposedly \textit{secure} configurations. This reflects a deeper problem in the \ac{ML} ecosystem: the slow adoption of newer library versions. Backward compatibility often takes precedence to preserve reproducibility, collaboration, and portability—sometimes at the silent cost of security, as our \acp{PoC} demonstrate.

Interestingly, while the Skops fix relies on obtaining explicit user confirmation and informing users of the associated security risks, the Keras team instead chose to extend the security measures of the new format to the legacy one, demonstrating a commitment to securing legacy versions.

To give a hint of how slow the adoption of newer versions is, we analyzed downloads after the release of Keras 3.9.0 (including critical security patches). Notably, older 2.x.x versions from 2023 still saw significantly higher download counts than most newer 3.x.x releases. More data are available in Appendix~\ref{appendix:adoption}. While this does not constitute a rigorous proof, it suggests a significant inertia within the ecosystem, where users continue to rely on older versions despite the availability of known (security) improvements. Consequently, developers need to maintain support for legacy mechanisms, further complicating the design of robust and secure sharing APIs.

% \mypar{Legacy Version Adoption in Keras}
% To measure how slow the adoption of newer versions is, we analyzed PyPI download statistics of Keras versions over time, collected through Google Cloud Public Datasets and queried using BigQuery\footnote{Queries, raw data, and scripts used to generate the plot are available at this link \url{https://anonymous.4open.science/r/sharing-usenix-8136/version_adoption_keras/README.md}.}. This is not intended to be an exhaustive study, which would be outside the scope of this work. Figure~\ref{fig:keras-downloads} presents the most downloaded Keras versions between March 5, 2025 (the release date of version 3.9.0, which included several critical security patches), and March 26, 2025 (the day before the release of version 3.9.1). This period represents the most favorable window for the adoption of version 3.9.0.

% As expected, version 3.9.0 was the most downloaded release during this interval. However, surprisingly, it was followed by two versions from 2023: 2.15.0 (1.1 million downloads) and 2.13.1 (900,000 downloads). No other version in the 3.x.x series received more than 500{,}000 downloads during this period and is therefore not shown in the plot.

% While this does not constitute a rigorous proof, it suggests a significant inertia within the ecosystem, where users often continue to rely on older versions despite the availability of known (security) improvements. Consequently, developers need to maintain support for legacy mechanisms, further complicating the design of robust and secure sharing APIs.

\begin{marcosboxgreyclosed}
\mypar{Suggestions for the Community}
While recognizing the complexity of the problem, developers should, whenever possible, require insecure legacy options to be explicitly enabled by users and provide clear warnings about the associated risks. Users, in turn, should maintain heightened skepticism toward legacy formats, as frameworks often prioritize compatibility over security.
\end{marcosboxgreyclosed}

\subsection{Model Scanning as Malware Analysis}
\begin{marcosboxgreyopen}
\textbf{Derived From:} \hyperref[sah]{SA.H}, \hyperref[ms]{MS}
\end{marcosboxgreyopen}

Automatic model scanners integrated into model hubs, while providing additional protection and being potentially useful, inherit well-known limitations of traditional signature-based malware detectors~\cite{souri2018state}. Moreover, they often adopt a framework-centric design, supporting only specific frameworks and formats, and frequently duplicate safeguards already implemented within the frameworks themselves. For instance, Protect AI~\cite{ProtectAI} defines framework- and format-specific threats and checks models against them~\cite{ProtectAIThreats}, restricting detection to already formalized weaknesses.

Our evaluation of Hugging Face model scanners (\hyperref[ms]{MS}), in addition to finding false positives and negatives (see evaluation results for details), shows that Skops, despite being promoted by Hugging Face as a secure persistence format, is unsupported by any integrated scanner. 
% This highlights the fragmented and incomplete nature of the model-scanner landscape. 
This highlights the fragmented and incomplete nature of the current landscape of model scanners integrated into model hubs.
Finally, the labels produced by these scanners can be misleading when no match is found. For instance, if a format is not supported, JFrog~\cite{JFrog} returns ``not a model'' and Protect AI reports ``No issue''. Such outputs, especially if they come from a limited analysis, can create a false sense of security, such as in traditional antivirus software~\cite{Lau2023UsenixSec}.

\begin{marcosboxgreyclosed}
\mypar{Suggestions for the Community} Inspired by the malware analysis domain, scanners should be treated only as a first line of defense. A possible direction is the adoption of behavioral analysis to move beyond static checks. Finally, hubs should promote greater transparency in report labels.
\end{marcosboxgreyclosed}

\subsection{Trusting ``Safe'': A Risk in Itself}
\begin{marcosboxgreyopen}
\textbf{Derived from:}
\hyperref[saf]{SA.F}, \hyperref[sah]{SA.H}, \hyperref[keras_first]{KV.\ref*{keras_first}}, \hyperref[keras_second]{KV.\ref*{keras_second}}, \hyperref[keras_third]{KV.\ref*{keras_third}}, \hyperref[skops_first]{SV.\ref*{skops_first}}, \hyperref[skops_second]{SV.\ref*{skops_second}}, \hyperref[skops_third]{SV.\ref*{skops_third}}, \hyperref[UP1]{UP.\ref*{UP1}}, \hyperref[UP2]{UP.\ref*{UP2}}, \hyperref[UP3]{UP.\ref*{UP3}, \hyperref[ms]{MS}}
\end{marcosboxgreyopen}
What becomes evident from our analysis is that a complete and truly secure solution for model sharing does not exist. Every approach, whether from hubs or frameworks, requires compromises or shifts part of the responsibility elsewhere. 
% Even those explicitly designed with security in mind and claiming minimal compromise can still fail: they may contain bugs, lack critical checks, or rely on incorrect assumptions. 
This is, unfortunately, a well-known reality in computer science, and model sharing is no exception.

However, this reality often clashes with the prevailing narrative and common beliefs. Labeling a mode or a model (after scanning) as ``safe'' or ``secure'' is rarely consistent with the actual risks involved. Instead, we should refer to these as approaches that \textit{attempt} to provide security hardening against known exploitation paths, establish mechanisms for trusted types, or detect potential (known) threats in models prior to download. However, such efforts do not inherently make these approaches "secure" or even "safe". % to use the more popular but less rigorous term often found in model-sharing narratives.

While we acknowledge the complexity in designing APIs accessible to users with diverse backgrounds, such as those in the \ac{ML} community, we believe there is a need for clear and transparent communication. Users often lack either the knowledge or the willingness to critically assess labels like ``secure'' or ``safe.'' Such terminology is a significant oversimplification that fails to reflect the complex (and less optimistic) reality.

This kind of messaging has real consequences for users. As shown in our survey (\hyperref[UP2]{UP.\ref*{UP2}} and \hyperref[UP3]{UP.\ref*{UP3}}), users exhibit increased security confidence when features like the \texttt{safe\_mode} option are enabled or when hubs advertise built-in security scanning. While choosing methods with some degree of security hardening is certainly a positive step, it is by no means sufficient to ensure \textit{true} security. A critical observation is that, with \texttt{safe\_mode} activated, more than 90\% of survey participants who did not choose cybersecurity as their field of expertise—but who had prior experience in model sharing—did not consider arbitrary code execution to be a concern. This misplaced trust is, in itself, a serious security issue that requires a coordinated community effort to be addressed.

\begin{marcosboxgreyclosed}
\mypar{Suggestions for the Community}
% Avoid oversimplified security labels, which give users a false sense of security, and instead promote genuinely informed trust. If simplifications are necessary, they should be made cautiously and without undue optimism.
Oversimplified security labels can create misplaced trust. When simplifications are necessary, present them with care to align with the actual security guarantees.
\end{marcosboxgreyclosed}

%{\color{blue}When simplifications are necessary, they should refrain from conveying unproven optimism.}
\section{Related Works}
% % {\color{blue}To further support the claim of fragmented and immature security research on model sharing presented in the introduction (Section~\ref{sec:intro}), we reviewed the state of the art, analyzing existing works on the risks of traditional code sharing and the software supply chain~\cite{duan2021towards,ohm2020backstabber,ladisa2023sok,jiang2022empirical,casey2024large,shenao2025sok,meiklejohn2025machine}, as well as prior research on model code injection~\cite{hua2024malmodel,wang2021evilmodel,wang2022evilmodel2,zhao2024models} and hub security~\cite{jiang2023exploring,jiang2023empirical,zhao2024models,Jones2024What}. Moreover, we observe important signs of the topic’s relevance in recent studies (e.g., on TensorFlow API abuse~\cite{zhu2025my}) and in talks at major (non-academic) security conferences such as DEF CON~\cite{ParzianDefcon33}. Due to space constraints, we refer the interested reader to Appendix~\ref{appendix:related}.}
% {\color{blue}To further support the claim of fragmented and immature security research on model sharing presented in the introduction (Section~\ref{sec:intro}), we reviewed the state of the art and analyzed recent evidence of the topic’s relevance. Due to space constraints, we refer the interested reader to Appendix~\ref{appendix:related}.}

Being an established research area, \textit{generic software} supply chain security has been extensively studied, with several works~\cite{duan2021towards,ohm2020backstabber,ladisa2023sok} identifying and systematizing its risks. While some insights from these studies can be extended to the model-sharing context, recent research has gone a step further by specifically examining supply chain issues in this context. In particular, Meiklejohn et al.~\cite{meiklejohn2025machine} framed \ac{ML} models as supply chain artifacts and proposed cryptographic mitigations for the identified threats, whereas Wang et al.~\cite{shenao2025sok} mapped vulnerabilities across the \ac{LLM} supply chain, noting that only 56\% of them currently have available fixes.

Building on the observation that model hubs have become central to model distribution and therefore represent a potential vector for supply chain attacks, Jiang et al. conducted a series of empirical studies examining the security of model-sharing platforms. In their initial work, they compared model hubs to the \textit{generic software} supply chain, categorizing them based on their access model (open or gated) and identifying the associated risks~\cite{jiang2022empirical}. Subsequent research explored typosquatting and impersonation threats~\cite{jiang2023exploring} and analyzed model reuse practices on Hugging Face~\cite{jiang2023empirical}. Across these studies, Jiang et al. found insufficient provenance verification, weak dependency management, and inadequate documentation, ultimately concluding that trust in hub-shared models is often misplaced. 
Extending the understanding of risks and practices within model hubs, Jones et al.~\cite{Jones2024What} analyzed the Hugging Face ecosystem, uncovering high model turnover rates, a correlation between popularity and documentation quality, and challenges in model management and reproducibility.

In parallel, other researchers have explored specific attack vectors. In particular, Casey et al.~\cite{casey2024large} and Zhao et al.~\cite{zhao2024models} demonstrated the prevalence of insecure serialization methods (e.g., pickle) in Hugging Face models, which expose users to arbitrary code execution. Building on this, Zhao et al. conducted a large-scale measurement study of malicious code poisoning on Hugging Face, discovering multiple infected models and dataset scripts.

Focusing more on the latter threat of code poisoning, Hua et al.~\cite{hua2024malmodel} introduced MalModel, which embeds executable payloads directly into deep learning model weights, demonstrating the feasibility of concealing malware within models. Similarly, EvilModel~\cite{wang2021evilmodel} and EvilModel 2.0~\cite{wang2022evilmodel2} showed that entire malware binaries can be hidden within model parameters without degrading performance.

From a framework-level perspective, Zhu et al.~\cite{zhu2025my} introduced the TensorAbuse attack, which exploits legitimate TensorFlow APIs to perform unintended operations (e.g., file access and network messaging) during model inference. This work highlights a gap in the state of the art on the security of model sharing.
This gap was further underscored by Cyrus Parzian, who presented a talk at DEF CON 33 on the risks of model sharing, using as examples the security concerns associated with pickle deserialization and ONNX models shared in \texttt{.exe} format~\cite{ParzianDefcon33}.

\mypar{Gaps in the State of the Art}
Prior work on the security of model sharing has primarily examined supply chain risks, malicious code injection, and abuses of specific TensorFlow APIs. However, no study has systematically evaluated the security of model-sharing approaches across different ML frameworks and hubs, nor critically questioned the actual guarantees of sharing mechanisms that are claimed to be \textit{secure}.
Our paper closes this gap. % by taking a broad view of the ML model-sharing landscape. 
We assess the security posture of widely used frameworks, uncover 0-day vulnerabilities (subsequently disclosed and assigned CVEs) in mechanisms advertised as secure but designed to support self-contained model artifacts, and examine how security narratives shape user perception. At the hub level, while prior work has focused mainly on supply-chain–related risks, we empirically assess the effectiveness of hub-integrated scanners and labeling practices, questioning their effectiveness and how they influence users’ sense of security.

\section{Future Directions}
\label{appendix:future}

In line with our goal of deepening understanding of the security challenges in loading \ac{ML} models, we highlight several potential directions for future research.

Beyond the framework and hub levels examined in this work, a third dimension may involve third-party libraries not included in official framework documentation and thus excluded from our selection criteria. The popularity of some (e.g., safetensors~\cite{safetensors}) makes it worthwhile to examine how they align with our identified categories and what unique implications they introduce.
% Beyond the framework and hub levels examined in this work, a third dimension might be represented by third-party libraries not presented in official framework documentation, and thus excluded from our selection criteria. The popularity of some of these (e.g., safetensors~\cite{safetensors}) might make it interesting to investigate how they fit within the categories identified in our study, as well as the specific implications and peculiarities they introduce. 
% Additionally, it might be worth exploring how users approach these libraries, specifically whether they apply a different level of skepticism compared to \textit{official} methods.
Additionally, it would also be interesting to explore how users approach these libraries, specifically whether they apply a different level of skepticism compared to \textit{official} methods.

LLM-specific formats (e.g., GGUF, GPTQ, AWQ) introduce additional challenges. Indeed, the prohibitive size of these models shifts users’ focus from training to inference, although even inference is often infeasible on local machines. Moreover, pre-trained models are typically released only by a few major vendors that have the computational resources to train them. These factors not only redirect the focus of sharing solutions toward optimized formats designed for inference, which are not required to be well-suited for architectural modifications (thus making restricted solutions less limiting), but may also alter the overall threat model due to the different distribution of actors involved. 
% On the other hand, their reliance on native code once again highlights the need to revisit traditional software security concerns, particularly those related to memory integrity.

Automatic model scanners also represent a broad area for future research. While we discussed their limitations, future work could systematically assess their performance at scale, particularly against diverse adversarial techniques. Inspired by advances in malware detection, additional approaches (e.g., dynamic analysis) could be explored.

% From the perspective of narrative and user perception, it would be valuable to further investigate how different sources contribute to shaping users’ understanding of risks, both in absolute terms and in relative importance. This includes not only official documentation and naming conventions, but also community forums and discussion platforms. 

\section{Conclusions}
% In this work, we evaluated the security posture of \acl{ML} model sharing. Our analysis shows that some framework settings and hubs do not offer security, while the protection provided by those that advertise security is often partial or misplaced. Mechanisms may rely on shifting responsibility to users or other workflow components, and the assumption that formats promoted as \textit{secure} inherently prevent exploitation proves unfounded in practice. By uncovering vulnerabilities and assessing user perception, we highlight a critical gap between the security narrative and the actual guarantees provided. 
% Our takeaways show that there is no straightforward silver bullet, be it file formats, framework flags, or model scanning, and that due to trade-offs with usability, security will inevitably come with costs that the community must acknowledge.
% Strengthening the ecosystem requires clearer communication of risks, continuous hardening of frameworks and hubs, and a culture where users treat shared models with the same caution as executable code.

% In this work, we evaluated the security posture of \acl{ML} model sharing across frameworks and hubs. We found that protection is inconsistent: many mechanisms provide no safeguards, while others often shift responsibility to users or impose strong restrictions on flexibility. Approaches promoted as ``secure'' fail to reliably prevent exploitation. By uncovering 0-day vulnerabilities and analyzing user perceptions of these mechanisms, we expose a critical gap between the security narrative and the guarantees actually provided. 
In this work, we evaluated the security posture of \acl{ML} model sharing across frameworks and hubs. We found that protection is inconsistent: many mechanisms provide no safeguards, while others shift responsibility to users or impose strong restrictions on flexibility. Even those promoted as \textit{secure} fail to reliably prevent exploitation. By uncovering 0-day vulnerabilities and analyzing user perceptions of these mechanisms, we exposed a critical gap between the security narrative and reality.
Our takeaways show that there is no straightforward silver bullet. 
% Data-based formats and model scanners fall short of ensuring actual security, and framework or hub naming choices compromise awareness. 
Data-based formats and model scanners integrated into model hubs fall short of ensuring actual security, and framework or hub naming choices further compromise user awareness.
Security inevitably involves trade-offs: reducing risk often limits usability, while support for legacy formats silently reintroduces old vulnerabilities. Automatic scanning can help, but its coverage is uneven, and results are sometimes misleading. Above all, shared models must be treated as code, and loading untrusted artifacts carries the same risks as executing untrusted software.

\section*{Acknowledgments}
This work was partially supported by the project SERICS (PE00000014) under the NRRP MUR program funded by the EU - NGEU. The authors would like to thank the framework and hub maintainers and developers, and the security teams involved in the coordinated disclosure process for their responsiveness, constructive engagement, and commitment to improving the security of the \ac{ML} ecosystem.

\bibliographystyle{IEEEtran}
\bibliography{bibliography}
\section*{Ethics Considerations}

This research systematizes and empirically evaluates risks associated with loading \ac{ML} models across widely used frameworks and hubs. The stakeholders for this research include \ac{ML} practitioners and researchers who load pre-trained models; framework developers and maintainers; model hubs and package repositories; and security researchers. End users and society at large are only indirectly impacted.

% Our research is guided by the beneficence principle: by analyzing the security postures of sharing mechanisms and the fixes for the vulnerabilities we identified, our goal has been to strengthen \ac{ML} security and raise awareness of real-world risks, and not to enable offensive applications.

%To obtain the maximum security benefits for all relevant stakeholders while minimizing potential harm, we proceeded as follows. 
Each vulnerability listed in this work was disclosed and discussed with the relevant maintainers in accordance with their disclosure guidelines, and we collaborated with them to design and validate effective mitigations. Misuse potentials have been mitigated through coordinated disclosure, including delaying the public release of any potentially exploit-enabling details until a timeline coordinated with the relevant maintainers and security teams.

% Our technical security research did not involve human subjects nor directly affect end users. Most of the experiments were conducted on our own systems and software instances. 
For experiments involving the Hugging Face Hub, we obtained explicit permission to upload research models for security testing (\textit{``if this is for security and research purposes, we grant you permission to upload your models. Please let us know if you found anything interesting.''}). For the survey we conducted, participation has been totally voluntary by design, and participants were informed appropriately about the intended use of the results. Sensitive data risks related to our survey were minimized by conducting it anonymously, avoiding the collection of personally identifiable information, and adhering to applicable laws, platform terms, and community norms. %We believe that this approach comprehensively adheres to the guiding principles of respect for persons, law, and public interest.

% In accordance with the fairness principle, bias risks are reduced by framing the problem through research questions and providing takeaways and suggestions to strengthen the security of model-sharing ecosystems.

To ensure transparency, we privately shared the complete preprint version of this paper with all vendors affected by the discovered vulnerabilities or directly examined in our study, so that they were informed before any broader dissemination. This step extended beyond the vulnerabilities and findings already disclosed responsibly and involved Google, Keras, Hugging Face, and Skops. We engaged in constructive discussions when needed or upon request, while maintaining full independence in the research process.

%Our decision to publish reflects a considered judgment that, after taking into consideration all of the above, increasing the awareness of actual security risks of \ac{ML} model sharing outweighs potential harm.

\section*{LLM usage considerations}
LLMs were used for editorial purposes in this manuscript, and all outputs were inspected by the authors to ensure accuracy and originality.

\appendices

\section{Prior CVEs on Keras and Skops}
\label{appendix:cve}
% We review CVEs for Keras and Skops disclosed prior to our work as an insight into the security maturity of the two frameworks analyzed in Section~\ref{sec:whensecisnotsec}. 
We review the CVEs disclosed prior to our work concerning either Keras or Skops model-sharing methods.
Our search included both the official CVE database (\url{https://cve.org}), using keywords such as "keras" and "skops", and the GitHub Security Advisory pages of the respective projects. No filters were applied to the publication time frame.

Before the publication of the first CVE we identified (\hyperref[keras_first]{KV.\ref*{keras_first}}) for Keras, the most recent CVE related to its model persistence was \texttt{CVE-2024-3660}~\cite{CVE-2024-3660}. This vulnerability enabled arbitrary code execution during model loading in versions of Keras prior to 2.13, which was released in March 2023. The root cause was the absence of the \texttt{safe\_mode} flag (introduced only in Keras 2.13) and the unrestricted nature of the \texttt{Lambda} layers in the default model formats used at the time (HDF5), which allowed deserialization of arbitrary Python code~\cite{CERT-VU-253266}. The only other CVE related to model loading published before our work was \texttt{CVE-2021-37678}~\cite{CVE-2021-37678}, which affected a YAML-based format that is no longer applicable, as it was deprecated following that disclosure. Notably, no CVEs had been assigned to the \texttt{.keras} format or to \texttt{safe\_mode}, and Keras’s GitHub Security Advisory page listed no advisories.

Interestingly, analysis of Keras’s changelog and commit history reveals silent fixes for security-relevant issues involving \texttt{safe\_mode} and not accompanied by advisories or CVE assignments. One such case is commit \texttt{57c94f3}~\cite{keras-commit-57c94f30} from January 2025, which addresses the insecure use of \texttt{numpy.load} with \texttt{allow\_pickle=True} when loading weights from \texttt{.npz} files—a rarely used but supported alternative to HDF5. Although this issue was reported earlier on bug bounty platforms (e.g., Huntr)~\cite{huntr-keras-rce} in February 2024, it appears to have been rejected at the time.

Similarly, for Skops, we found only one prior CVE: \texttt{CVE-2024-37065}~\cite{CVE-2024-37065}. This vulnerability allowed arbitrary code execution when using the now-deprecated \texttt{trusted=True} flag. Importantly, this setting was used internally within parts of the Skops codebase, such as the update CLI tool, potentially exposing users even if they did not explicitly enable it~\cite{SchulzSkopsAdvisory}. Beyond this, we found no other vulnerabilities or advisories disclosed publicly.
\begin{figure}[t]
\centering
\includegraphics[width=\linewidth]{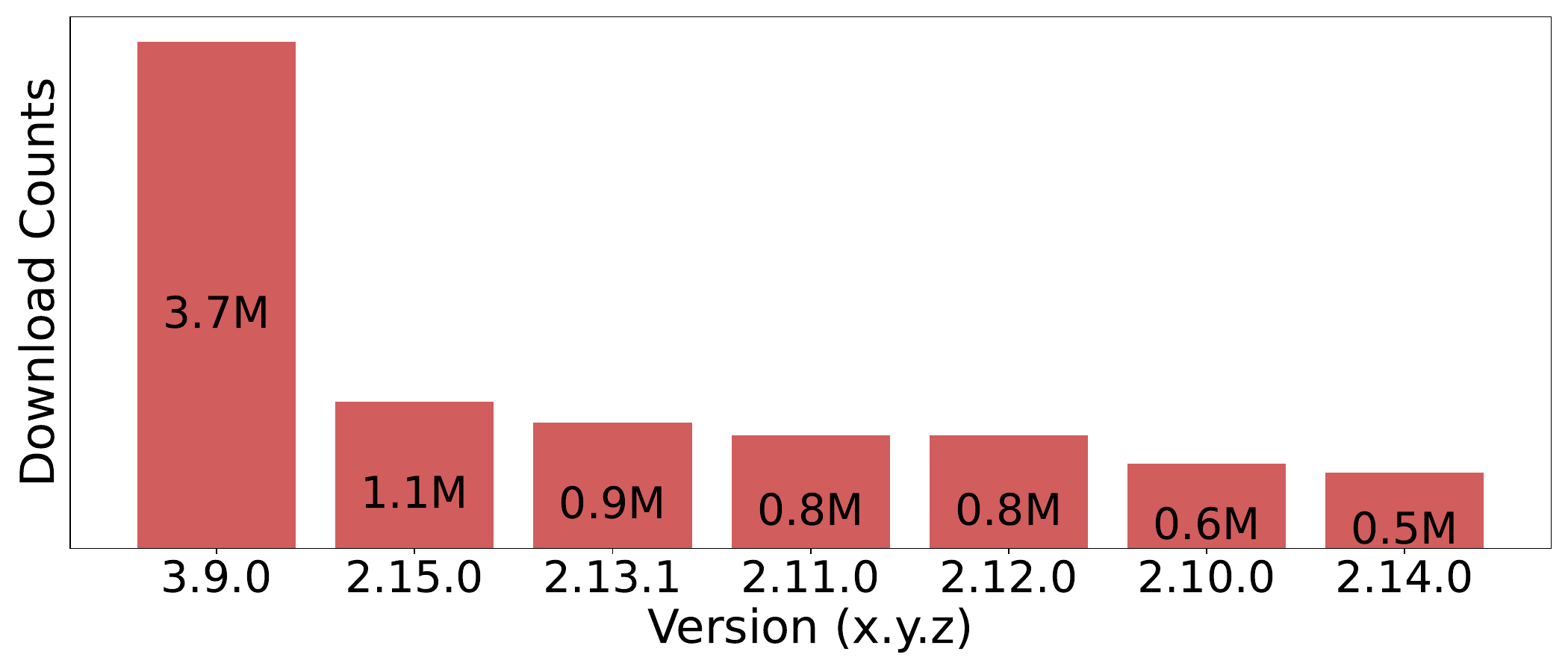}
\caption{Download statistics of Keras versions between March 5, 2025, and March 26, 2025 (only versions with $>$500k downloads).}
% \caption{Keras download statistics (Mar 5–26, 2025; versions with $>$500k downloads).}
\label{fig:keras-downloads}
\end{figure}

\section{Legacy Version Adoption Rate in Keras}
\label{appendix:adoption}
We analyzed PyPI download statistics for Keras over time, following the release of version 3.9.0, which included critical security fixes. 
While this is not intended to be an exhaustive study, which would be outside the scope of this work, our intent is to give a hint at the adoption rate of newer versions.
The statistics were collected from Google Cloud Public Datasets and queried using BigQuery.
We made the queries, raw data, and scripts used to generate the plot publicly available.

Figure~\ref{fig:keras-downloads} presents the most downloaded Keras versions between March 5, 2025 (the release date of version 3.9.0), and March 26, 2025 (the day before the release of version 3.9.1). This represents the most favorable time window for the adoption of version 3.9.0, which was, as expected, the most downloaded release. However, surprisingly, it was followed by two versions from 2023: 2.15.0 (1.1 million downloads) and 2.13.1 (900,000 downloads). No other version in the 3.x.x series received more than 500{,}000 downloads during this period and is therefore not shown in the plot.

\begin{listing}[t]
\centering
% \begin{minted}[fontsize=\scriptsize, frame=none, bgcolor=gray!5]{json}
\begin{minted}[fontsize=\footnotesize, frame=none, bgcolor=gray!5]{json}
{
   "module": "subprocess",
   "class_name": "run",
   "inbound_nodes": [
      {
         "args":[
            "/bin/sh"
         ],
         "kwargs":{
            
         }
      }
   ]
}
\end{minted}
\caption{Simplified malicious \texttt{config.json} snippet for \hyperref[keras_first]{KV.\ref*{keras_first}}. Keras interprets \texttt{subprocess.run} as a model layer. Arguments are passed via \texttt{inbound\_nodes}, which define input–output relations within Keras’s model computation graph.}
\label{fig:malicious-config}
\end{listing}
\begin{listing}[t]
\centering
\begin{minipage}{\linewidth}
\begin{minted}[
    fontsize=\footnotesize,
    frame=none,
    bgcolor=gray!5,
    breaklines=true,
    breakanywhere=true
]{json}
{
   "module":"keras.layers",
   "class_name": "Lambda",
   "config":{
      "name": "set_global_state",
      "function":{
         "module": "keras.src.backend.common.global_state",
         "class_name": "function",
         "config": "set_global_attribute",
         "registered_name": "function"
      },
      "arguments":{
         "value": false
      }
   },
   "name": "set_global_state",
   "inbound_nodes":[
      {
         "args":[
         ],
         "kwargs":{
            "inputs": "safe_mode_saving"
         }
      }
   ]
}
\end{minted}
\caption{Partial malicious \texttt{config.json} snippet for \hyperref[keras_second]{KV.\ref*{keras_second}}. A \texttt{Lambda} layer is abused to disable safe mode via \texttt{set\_global\_attribute("safe\_mode\_saving", value=False)}. Arguments are passed through both the top-level \texttt{inbound\_nodes} key and the \texttt{arguments} key within the \texttt{Lambda} layer configuration.}
\label{fig:disable-safe-mode}
\end{minipage}
\end{listing}
\begin{listing}[t]
\centering
\begin{minipage}{\linewidth}
\begin{minted}[
    fontsize=\footnotesize,
    frame=none,
    bgcolor=gray!5,
    breaklines=true,
    breakanywhere=true
]{json}
{
   "__class__": "int",
   "__module__": "builtins",
   "__loader__": "MethodNode",
   "content":{
      "obj":{
         "__class__": "int",
         "__module__": "builtins",
         "__loader__": "MethodNode",
         "content":{
            "obj":{
               "__class__": "QuadraticDiscriminantAnalysis",
               "__module__": "sklearn.discriminant_analysis",
               "__loader__": "ObjectNode",
               "__id__": 1
            },
            "func": "decision_function"
         }
      },
      "func": "__builtins__"
   }
}
\end{minted}
\caption{Malicious \texttt{schema.json} snippet for \hyperref[skops_first]{SV.\ref*{skops_first}}. \texttt{QuadraticDiscriminantAnalysis} from Scikit-learn (trusted by default) is instantiated using an \texttt{ObjectNode}. Then, a first \texttt{MethodNode} accesses its \texttt{decision\_function} method and a second \texttt{MethodNode} retrieves the \texttt{\_\_builtins\_\_} dictionary, bypassing Skops’ checks.}
\label{fig:methodnode-chain}
\end{minipage}
\end{listing}
\begin{listing}[t]
\centering
\begin{minipage}{\linewidth}
\begin{minted}[
    fontsize=\footnotesize,
    frame=none,
    bgcolor=gray!5,
    breaklines=true,
    breakanywhere=true
]{json}
{
   "__class__": "call",
   "__module__": "sklearn.SGDRegressor",
   "__loader__": "OperatorFuncNode"
}
\end{minted}
\caption{Malicious \texttt{schema.json} snippet for \hyperref[skops_second]{SV.\ref*{skops_second}}. The validated type string is \texttt{sklearn.SGDRegressor.call}, which may appear benign and related to the target model, but what is actually invoked is \texttt{operator.call}.}
\label{fig:operatorfuncnode}
\end{minipage}
\end{listing}

\FloatBarrier
\null
\null
\null
\null
\null
\null
\null
\null
\null
\null
\null
\null
\null
\null
\null
\null
\section{Meta-Review}
The following meta-review was prepared by the program committee for the 2026
IEEE Symposium on Security and Privacy (S\&P) as part of the review process as
detailed in the call for papers.

\subsection{Summary}
The paper investigates supply-chain attacks targeting ML model-loading workflows. It considers a scenario in which an adversary maliciously manipulates a model to trigger arbitrary code execution when it is loaded by a victim. The paper surveys security mechanisms for model loading in popular ML frameworks and model hubs, conducts a manual vulnerability assessment of security-oriented features, and evaluates user perceptions of model-loading safety. The analysis uncovers multiple vulnerabilities in Keras's safe\_mode and Hugging Face's model scanning mechanisms.

\subsection{Scientific Contributions}
\begin{itemize}
\item Identifies an Impactful Vulnerability.
\item Provides a Valuable Step Forward in an Established Field.
\end{itemize}

\subsection{Reasons for Acceptance}
\begin{enumerate}
\item The paper provides a systematic survey that highlights the fragmentation and lack of standardization in current model-sharing practices.
\item It identifies multiple exploits in secure loading modes, raising serious concerns about the effectiveness of existing safeguards in model-sharing infrastructures.
\end{enumerate}

% \subsection{Noteworthy Concerns} % Exclude if your meta-review does not have noteworthy concerns
% None

% \section{Response to the Meta-Review} % Optional

% Less than 500 words response to the meta-review. The response to the
% meta-review is optional. Provide a response if you disagree with the
% meta-review. Shepherds will only deny responses to meta-reviews if they are too
% long or are abusive / inappropriate.

% % use section* for acknowledgment
% \ifCLASSOPTIONcompsoc
%   % The Computer Society usually uses the plural form
%   \section*{Acknowledgments}
% \else
%   % regular IEEE prefers the singular form
% \fi

\end{document}